%% file: paper.tex
\documentclass[prb,twocolumn,superscriptaddress,showpacs]{revtex4}
\usepackage{graphicx,amsmath,amssymb,bm}

\newcommand{\ket}[1]{|#1\rangle}

\begin{document}

\title{{\it Ab initio} computation of circular quantum dots}

\author{M.~Pedersen Lohne}

\affiliation{Department of Physics, University of Oslo, N-0316 Oslo, Norway}

\author{G.~Hagen}

\affiliation{Physics Division, Oak Ridge National Laboratory, Oak Ridge, TN 37831, USA}

\author{M.~Hjorth-Jensen}

\affiliation{Department of Physics and Center of Mathematics for
  Applications, University of Oslo, N-0316 Oslo, Norway}

\author{S.~Kvaal}

\affiliation{Center of Mathematics for Applications, University of Oslo, N-0316 Oslo, Norway}

\author{F.~Pederiva}

\affiliation{Dipartimento di Fisica, Universit\`a di Trento, and I.N.F.N., Gruppo Collegato di Trento, I-38123 Povo, Trento, Italy}
\

\begin{abstract}
  We perform coupled-cluster and diffusion Monte Carlo calculations of the energies of circular quantum dots  up to $20$ electrons.  The coupled-cluster calculations include triples corrections and   a renormalized Coulomb interaction
defined for a given number of low-lying oscillator shells. Using such a renormalized Coulomb interaction brings the coupled-cluster calculations with triples correlations in excellent agreement with the diffusion Monte Carlo calculations.
This opens up perspectives for doing  {\em ab initio} calculations for much larger systems of electrons. 
\end{abstract}


\maketitle 

\section{Introduction}\label{sec:intro}

Strongly confined electrons
offer a wide variety of complex and subtle phenomena which pose severe 
challenges to existing many-body methods.
Quantum dots in particular, that is, electrons confined in semiconducting heterostructures,
exhibit, due to their small size, discrete quantum levels. 
The ground states of, for example, circular dots
show similar shell structures and magic numbers 
as seen for atoms and nuclei. These structures are particularly evident in
measurements of the change in electrochemical potential due to the addition of
one extra electron, 
$\Delta_N=\mu(N+1)-\mu(N)$. Here $N$ is the number of electrons in the quantum dot, and
$\mu(N)=E(N)-E(N-1)$ is the electrochemical potential of the system.
Theoretical predictions of $\Delta_N$ and the excitation energy spectrum require
accurate calculations of ground-state and of excited-state energies.

The above-mentioned quantum mechanical levels can, in turn, be tuned by means
of, for example, the application of various external fields.  
The spins of the electrons in quantum dots
provide a natural basis for representing so-called qubits \cite{divincenzo1996}. The capability to manipulate
and study such states is evidenced by several recent experiments (see, for example, Refs.~\onlinecite{exp1,exp2}).
Coupled quantum dots are particularly interesting since so-called  
two-qubit quantum gates can be realized by manipulating the 
exchange coupling which originates from the repulsive Coulumb interaction 
and the underlying Pauli principle.  For such states, the exchange coupling splits singlet and triplet states, 
and depending on the shape of the confining potential and the applied magnetic field, one can allow
for electrical or magnetic control of the exchange coupling. In particular, several recent experiments and 
theoretical investigations have analyzed the role of effective spin-orbit interactions 
in quantum dots \cite{exp5,exp6,pederiva2010,spinorbit} and their influence on the exchange coupling.

A proper theoretical understanding of the exchange coupling, correlation energies, 
ground state energies of quantum dots, the role of spin-orbit interactions
and other properties of quantum dots as well, requires the development of appropriate and reliable  
theoretical  few- and many-body methods. 
Furthermore, for quantum dots with more than two electrons and/or specific values of the 
external fields, this implies the development of few- and many-body methods where   
uncertainty
quantifications are provided.  
For most methods, this means providing an estimate of the error due 
to the truncation made in the single-particle basis and the truncation made in 
limiting the number of possible excitations.
For systems with more than three or four electrons,  {\em ab initio} methods that have been employed in studies of quantum dots are
 variational and diffusion Monte Carlo \cite{harju2005,pederiva2001, pederiva2003}, path integral approaches \cite{pi1999}, large-scale diagonalization (full configuration 
interaction) \cite{Eto97,Maksym90,simen2008,modena2000}, and to a very limited extent 
coupled-cluster theory \cite{bartlett2007,bartlett2003,indians}. 
Exact diagonalization studies are accurate for a very small number
of electrons, but the number of basis functions needed to obtain a given
accuracy and the computational cost grow very rapidly with electron number.
In practice they have been used for up to eight electrons\cite{Eto97,Maksym90,modena2000}, but the accuracy is
very limited for all except $N\le 3$ (see, for example, Refs.~\onlinecite{simen2008,kvaal2009}).  
Monte Carlo methods have been applied up to $N=24$ electrons 
\cite{pederiva2001,pederiva2003}. Diffusion Monte Carlo, with statistical and systematic errors, provide, in principle,
exact benchmark solutions to various properties of quantum dots. However, 
the computations start becoming rather time-consuming for larger systems.   
Hartree\cite{Kum90}, restricted Hartree-Fock, spin- and/or space-unrestricted
Hartree-Fock\cite{Fuj96,Mul96,Yan99} and
local spin-density, and current density functional methods\cite{Kos97,Hir99,finns1,finns2}
give results that are satisfactory for a qualitative understanding of some
systematic properties. However, comparisons with exact results show
discrepancies in the energies that are substantial
on the scale of energy differences. 

Another many-body method with the potential of providing reliable error estimates and accurate 
results is coupled-cluster theory, with its various levels of truncations. Coupled-cluster theory is the 
method of choice in quantum chemistry, 
atomic and molecular physics \cite{bartlett2007,helgaker2003}, and has recently been applied 
with great success in nuclear physics as well (see, for 
example, Refs.~\onlinecite{hagen2008,hagen2009,hagen2010a,hagen2010b}). 
In nuclear physics, with our spherical basis codes, 
we expect now to be able to perform {\em ab initio} calculations of nuclei
up to  $^{132}$Sn with more than 20 major oscillator shells. The latter implies dimensionalities
of more than $10^{100}$ basis Slater determinants, well beyond the reach of the full configuration
interaction approach. Coupled-cluster theory
offers a many-body formalism which allows for systematic expansions and error estimates
in terms of truncations in the basis of single-particle states \cite{schneider2008}. 
The cost of the calculations scale gently with the number of particles and single-particle states, and we expect
to be able to study quantum dots up to 50 electrons without a spherical symmetry.
The main advantage of the coupled-cluster  method over, say, full configuration approaches relies on the fact that it offers an attractive truncation scheme at a much lower computational cost. It preserves, at the same time, important features such as size extensivity.    

The aim of this work is to apply coupled-cluster theory with the
inclusion of triples excitations through the highly accurate and
efficient $\Lambda$-CCSD(T)  approach \cite{Taube08,taube2} for 
circular quantum dots up to $N=20$ electrons, employing different strengths of the applied magnetic field. 
The results from these calculations are compared in turn with, in principle, exact diffusion Monte Carlo 
calculations.  Moreover, this work introduces a technique widely applied in the nuclear many-body problem, 
namely that of a renormalized two-body Coulomb interaction. Instead of using the free Coulomb 
interaction in an oscillator basis,
we diagonalize the two-electron problem exactly using a tailor-made
basis in the centre-of-mass frame.\cite{simen2008}
The obtained eigenvectors and eigenvalues are used, in turn, to obtain, via a similarity transformation, an effective 
interaction defined for the lowest $10-20$ oscillator shells. These shells define our effective Hilbert space
where the coupled-cluster calculations are performed. This technique has been used with great success in the nuclear many-body problem, in particular since the strong repulsion at short interparticle distances of the nuclear 
interactions requires a renormalization of the short-range part \cite{navratil1,navratil2}. 
With this renormalized Coulomb interaction and coupled-cluster
calculations with triples excitations included through the $\Lambda$-CCSD(T)
approach, we obtain results in close agreement with 
the diffusion Monte Carlo calculations.  This opens up many interesting avenues for {\em ab initio} studies
of quantum dots, in particular for  systems beyond the simple circular quantum dots.  

This article is organized as follows.  Section II introduces
(i) the Hamiltonian and interaction for circular quantum dots, 
(ii) the basic ingredients for obtaining an effective interaction using a 
similarity-transformed Coulomb interaction, then (iii) a brief 
review of coupled-cluster theory and the $\Lambda$-CCSD(T) approach, 
and finally (iv) the corresponding details behind the diffusion Monte Carlo calculations.
In Section III, we present our results, whereas Section IV is devoted to our conclusions and
perspectives for future work.

\section{Coupled-cluster theory and diffusion Monte Carlo}\label{sec:methods}
In this section we present first our Hamiltonian in Subsection \ref{subsec:physics}; thereafter we discuss how to 
obtain a renormalized two-body interaction in an effective Hilbert space. In Subsection  \ref{subsec:cctriples}
we present our coupled-cluster approach, and finally in Subsection \ref{subsec:dmc}  we briefly review our
diffusion Monte Carlo approach.

\subsection{Physical systems and model Hamiltonian}\label{subsec:physics}

We will assume that our problem can be described entirely by  a non-relativistic 
many-electron Hamiltonian $\hat{H}$, resulting in the Schr\"odinger equation
\begin{equation}
\label{eq: schrödinger equation}
\hat{H}|\Psi\rangle = E|\Psi\rangle,
\end{equation}
with $|\Psi\rangle$ being the eigenstate and $E$ the eigenvalue.
The many-electron Hamiltonian is normally written in terms of a non-interacting part $\hat{H}_0$ and 
and interacting part $\hat{V}$, namely
\[
\hat{H} = \hat{H}_0 + \hat{V}= \sum_{i=1}^N \hat{h}_i + \sum_{i<j}^N \hat{v}_{ij}, 
\]
where $\hat{H}_0$ is the (one-body) Hamiltonian of the non-interacting system, 
and $\hat{V}$ denotes the (two-body) Coulomb interaction. In general, 
the Schr\"odinger equation (\ref{eq: schrödinger equation}) cannot be 
solved exactly. 

We define the reference Slater determinant $|\Phi_0\rangle$ 
as the ground state of the non-interacting 
system by filling all the lowest-lying single-particle orbits. Since we will limit ourselves to 
systems with filled shells, this may be a good approximation, in particular if the single-particle field
is the dominating contribution to the total energy. The non-interacting Schr\"odinger equation reads
\begin{equation}
\label{eq: non-interacting schrödinger equation}
\hat{H}_0|\Phi\rangle = e_0|\Phi_0\rangle,
\end{equation}
where
\[
\hat{H}_0 = \sum_{i=1}^N \hat{h}_i = \sum_{i=1}^N\left[\hat{t}_i + \hat{v}_{\mathrm{con}}({\bf r}_i)\right].
\]
The terms $\hat{t}_i$ and $\hat{v}_{\mathrm{con}}({\bf r}_i)$ are the kinetic-energy operator and the confining potential 
(from an external applied potential field) of electron $i$, respectively. The vector ${\bf r}_i$ represents the position in two dimensions of electron $i$.
Due to the 
identical and fermionic nature of electrons, the eigenstates of 
Eq.~(\ref{eq: non-interacting schrödinger equation}) are Slater determinants, with the general form
\[
|\Phi\rangle = |ijk\dots m\rangle = \hat{a}^{\dagger}_{i}a^{\dagger}_{j}\hat{a}^{\dagger}_{k}..\hat{a}^{\dagger}_{m}|0\rangle,
\]
with $\hat{a}^{\dagger}$ being standard fermion creation operators (and $\hat{a}$ being annihilation operators).
The single-particle eigenstates $|i\rangle=\hat{a}^{\dagger}_{i}|0\rangle$ 
and eigenenergies $\varepsilon_i$ are given by the solutions of the
one-particle Schr\"odinger equation governed by the operator $\hat{h}_i$.
Since the total energy of the non-interacting system is given by the 
sum of single-particle energies $\epsilon_i$, we have
\[
 e_0= \sum_{i=1}^N\varepsilon_i,
\]
the reference determinant $|\Phi_0\rangle$ is obviously the Slater determinant 
constructed from those orbitals with single-particle energies that yield the 
lowest total energy. In the particle-hole formalism, orbitals in the occupied 
space are referred to as hole states, while orbitals in the virtual space are 
denoted particle states. In principle, any complete and orthogonal
single-particle basis can be used. However, since our coupled-cluster
approach involves solution of a set of non-linear equations, 
it is preferable to start from a basis that produces a mean-field
solution not too far away from the ``exact'' and fully correlated many-body solution. 
Therefore our main results will be obtained using the Hartree-Fock
basis as a starting point for our coupled-cluster calculations. The Hartree-Fock basis is obtained
from a linear expansion of harmonic oscillator basis functions, such
that the expectation value of the Hamiltonian is minimized. 

For the diffusion Monte Carlo calculations, it is also necessary to start from
a model wavefunction that is used as importance function in the sampling, as
we will discuss later.
The Slater determinant part, in this case, is built starting from the self-consistent orbitals
generated in a Local Density Approximation  calculation, in order to include
as much information as possible about both exchange and correlation effects at the
one-body level. Explicit two-body correlations are then included  as an elaborate
Jastrow factor; see Subsection \ref{subsec:dmc} for further details.   

Our  model Hamiltonian~\cite{Ash96} for a quantum dot consists of 
a two-dimensional system of $N$ electrons moving in the $z=0$ plane,
confined by a parabolic lateral confining potential $V_{\rm con}({\bf r})$.
The  Hamiltonian is
\[
\hat{H}=\sum_{i=1}^N(-{\frac{\hbar^2}{2m_{e}m^*}}\nabla_i^2+ V_{\rm con}({\bf r}_i)) +
\frac{e^2}{\epsilon}\sum_{i<j}^N\frac{1}{|{\bf r}_i-{\bf r}_j|}.
\]
In the above equation, $m^*$ is a parameter relating the bare electron mass to an effective mass,
and $\epsilon$ is the dielectric constant of the semiconductor.
In  the following (if not explicitly specified otherwise), we
will use effective atomic units, defined by $\hbar=e^2/\epsilon=m_{e}m^*=1$. In this
system of units, the length unit is the Bohr radius $a_0$ times $\epsilon/m^*$,
and the energy has units of Hartrees times $m^*/\epsilon^2$.
As an example, for the GaAs quantum dots, typical values are 
$\epsilon=12.4$ and $m^*=0.067$. The effective
Bohr radius $a_0^*$ and effective Hartree H$^*$ are
$\simeq 97.93 $ \AA\  and $\simeq 11.86$ meV,
respectively.
In this work we will consider
circular dots only with $N=2$, $N=6$, $N=12$, and $N=20$ electrons confined by a
 parabolic potential $V_{\rm con}({\bf r})=m_{e}m^* \omega^2r^2/2$. The numbers $N=2$, $N=6$, $N=12$, and $N=20$
are so-called magic numbers corresponding to systems with closed harmonic oscillator shells, and hopefully a single-reference Slater determinant yields a good starting 
point  for our calculations.

The one-body part of our Hamiltonian becomes 
\[
\hat{H}_0=\sum_{i=1}^N\left(-{\frac{1}{2}}\nabla^2_{i}+\frac{ \omega^2}{2}r^2_{i} \right),
\]
whereas the interacting part is
\[
\hat{V}=\sum_{i<j}^N\frac{1}{|{\bf r}_i-{\bf r}_j|}.
\]
The unperturbed part of the Hamiltonian yields the  single-particle energies
\begin{equation}
\epsilon_i = \omega\left(2n+|m| + 1\right),
\end{equation}
where $n = 0,1,2,3,..$ and $m = 0, \pm 1, \pm 2,..$. The index $i$ runs from $0,1,2,\dots$.
The shell-structure 
is clearly deduced from this expression. We define $R$ as the shell index. We 
will denote the shell with the lowest energy as $R=1$, the shell with the second 
lowest energy as $R=2$, and so forth. Hence, 
\begin{equation}
R_i \equiv \frac{\epsilon_{i-1}}{\omega} \hspace{1cm} (i = 1,2,3,...).
\end{equation} 
In the calculations, we limit ourselves to values of $\omega=0.28$ a.u.~(atomic units), $\omega=0.5$ a.u., and $\omega=1.0$ a.u. For higher values of the oscillator frequency, the contribution to the energy from 
the single-particle part dominates
over the correlation part. The value $\omega=1.0$ is an intermediate 
case which also allows for comparison
with Taut's exact solution for $N=2$, see Ref.~\onlinecite{taut1994}, while $\omega=0.5$ and $0.28$   represent cases
where correlations are stronger, due to the lower average electron density in the dot.

\subsection{Effective interaction}\label{subsec:effint}

Whenever a single-particle basis is introduced in order to carry out a
many-body calculation, it must be truncated. The harmonic oscillator basis is
the \emph{de facto} standard for quantum dots and nuclear structure calculations. In
nuclei, the intrinsic Hamiltonian is most easily treated using this basis, and
for quantum dots the confining potential is to a good approximation harmonic.

However, the discrete Hilbert space $\mathcal{H}$ obtained from such a truncation grows
exponentially with the number of particles. For example, allowing $n$
single-particle states and $N$ particles,
\begin{equation*}
 \dim(\mathcal{H}) = \begin{pmatrix} n \\ N \end{pmatrix}.
\end{equation*}
As an example, if we distribute $N=6$ electrons in the total number of
single-particle states defined by 10 major oscillator shells, we have $n=110$,
resulting in $\dim(\mathcal{H}) \approx 2.3 \times 10^{13}$ Slater
determinants.  This number is already beyond the limit of present full
configuration interaction approaches.  In our coupled-cluster calculations we
perform studies up to some $20$ major shells. For $20$ shells, the total
number of single-particle states is $n=420$, for which $\dim(\mathcal{H})
\approx 1.3\times 10^{18}$, well beyond reach of standard diagonalization
methods in the foreseeable future. For $20$ electrons in $20$ shells, the
number of Slater determinants is much larger, $7.6\times 10^{33}$ in total.

But even if we could run large configuration interaction calculations, the
convergence of the computed energies as a function of the chosen
single-particle basis is slow for a harmonic oscillator basis, mainly due to
the fact that this basis does not properly take into
account the cusp condition at $|{\bf r}_i-{\bf r}_j|=0$ of the Coulomb
interaction.  In fact, the error $\Delta E$ in the energy for a quantum dot
problem, when increasing the dimensionality to one further shell with a
harmonic oscillator basis, behaves like
\begin{equation}
\Delta E \sim O(R_{ho}^{-k+\delta-1}).\label{eq:simenproof} 
\end{equation}
Here, $k$ is the number of times a given wave function $\Psi$ may be
differentiated weakly, $\delta\in[0,1)$ is a constant and $R_{ho}$ is the last
oscillator shell. The derivation of the latter relation is
detailed in Ref.~\onlinecite{kvaal2009}, together with extensive discussions of the
convergence properties of quantum dot systems.  For the ground state of the
two-electron quantum dot, we have precisely $k=1$, while for higher electron
numbers one observes $k = O(1)$. This kind of estimate tells us that an approximation using
only a few HO eigenfunctions necessarily will give an error depending directly
on the smoothness $k$.

Although the coupled-cluster method allows for the inclusion of much larger
single-particle spaces, the slow convergence of the energy seen in full
configuration interaction calculations applies to this method as well as it
approximates the configuration-interaction solution using the same set of
single-particle functions.  For an overview of coupled cluster error analysis,
see Refs.~\onlinecite{schneider2008,kutzelnigg1991}.

One way to circumvent the dimensionality problem is to introduce a
renormalized Coulomb interaction $\hat{V}_\text{eff}$ defined for a limited number of low-lying
oscillator shells. Such techniques have been widely used in nuclear many-body
problems (see, for example,
Refs.~\onlinecite{simen2008,mhj1995,navratil2009}). For quantum dots, this was
first applied to a configuration interaction calculation by Navratil {\em et
  al.} \cite{navratildots}, albeit for a different quantum dot model. But the
potential of this method has not been explored fully, except for recent
prelimary studies in Refs.~\onlinecite{simen2008,kvaal2009,kvaal2007}, which
demonstrate a significant improvement of the eigenvalues.  Furthermore, we
expect that the potential of this method is of even greater interest when
linked up with an efficient many-body method like the coupled-cluster
approach.

The recipe for obtaining such an effective interaction is detailed in several
works (see, for example, Refs.~\onlinecite{klein1974,simen2008,navratil2009}). Here we
give only a brief overview.

The Hilbert space $\mathcal{H}$ is divided into two parts $P\mathcal{H}$ and
$Q\mathcal{H}$, where $P$ and is the orthogonal projector onto the smaller,
effective model space, and $Q=1-P$. Note here, that $P\mathcal{H}$ will be the
space in which we do our many-body computations, and $\mathcal{H}$ is, in
principle, the {\em whole untruncated} Hilbert space. The interaction operator
$\hat{V}$ is considered a perturbation, and we introduce a convenient
complex parameter $z$ and study $\hat{H}(z) = \hat{H}_0 + z\hat{V}$. Setting
$z=1$ recovers the original Hamiltonian.

Consider a similarity transformation of $\hat{H}(z)$ defined by
\begin{equation}
 \tilde{H}(z) \equiv e^{-X(z)} \hat{H}(z) e^{X(z)}
\label{eq:simtrafo}
\end{equation}
where the operator $X(z)$ is such that the property
\begin{equation}
  Q\tilde{H}(z) P = 0
\label{eq:simtrafo2}
\end{equation}
holds. Equation \eqref{eq:simtrafo} must not be confused with equations from
coupled-cluster theory. The idea is that $X(z)$ should be determined from
perturbation theory, which gives an analytic operator function with $X(0) = 0$.

The most important consequences of these equations are that (i)
$\tilde{H}$ have identical eigenvalues with $\hat{H}$, (ii) that
there are $D=\dim(P\mathcal{H})$ eigenvalues whose eigenvetors are {\em entirely
  in the model space $P\mathcal{H}$}. Thus, the effective Hamiltonian defined by 
\begin{equation}
 \hat{H}_\text{eff}(z) \equiv P\tilde{H}(z)P
\end{equation}
is a model-space operator with $D$ exact eigenvalues. At $z=0$ these are the
unperturbed eigenvalues, and these are continued analytically as $z$ approaces
$z=1$. 

Equations \eqref{eq:simtrafo} and \eqref{eq:simtrafo2} are not sufficient to determine
$X(z)$ uniquely. The order-by order-expansion of $X(z)$ must be supplied with side
conditions. One of the most popular conditions is that $X(z)^\dag = -X(z)$ such
that $\tilde{H}(z)$ is Hermitian, and additionally that the effective
eigenvectors are \emph{as close as possible to the exact eigenvectors}, i.e.,
that the quantity $\Delta$ defined by
\begin{equation}
\Delta \equiv \sum_{k=1}^D \| \ket{\Psi_k} - \ket{\Psi_{\text{eff},k}} \|^2
\end{equation}
is minimized, where $\ket{\Psi_{\text{eff},k}}$ are the eigenvectors of
$\hat{H}_\text{eff}$, see Ref.~\onlinecite{simen2008}. One can obtain a formula for $X(z)$ in this case, namely
\begin{equation*}
 X = \operatorname{artanh}(\omega-\omega^\dag),
\end{equation*}
where $\omega=Q\omega P$ is the operator such that $\exp(\omega) P\ket{\Psi_k} =
\ket{\Psi_k}$. 

Order-by-order expansion of $\tilde{H}(z)$ reveals that it contains $m$-body
terms for all $m\leq N$, even though $\hat{V}$ only contains two-body
interactions. However, the many-body terms can be shown to be of lower
order\cite{klein1974} in $z$. By truncating $\tilde{H}(z)$ at terms at the
two-body level we obtain the so-called {\em sub-cluster approximation} to the
effective Hamiltonian. This can be computed by exact diagonalization of the
two-body problem; a simple task for the quantum dot problem
\cite{simen2008b}.

The one-body part of $\tilde{H}$ is always $H_0$, so it is natural to define
the effective {\em interaction} by
\begin{equation}
 H_\text{eff} = H_0 + V_{\text{eff}}.
\end{equation}

The reader should, however, keep in mind that the sub-cluster approximation
always produces missing many-body correlations when inserted in a many-body
context. The size of this source of error can only be quantified {\em a
  posteriori}, either by comparison with experiment and/or exact calculations
(see, for example, Ref.~\onlinecite{hdhk2010}) for a discussion on missing
many-body physics and the nuclear many-body problem.

\subsection{Coupled-cluster method}\label{subsec:cctriples}

The single-reference coupled-cluster theory
is based on the exponential ansatz for the
ground-state wave function of the $N$-electron system,
\[
|\Psi_{0}\rangle = e^{T} |\Phi_0\rangle,
\]
where $T$ is the cluster operator (an $N$-particle-$N$-hole excitation operator)
and $|\Phi_0\rangle$ is the corresponding reference determinant
(defining our chosen closed-shell system or vacuum) obtained
by performing some mean-field calculation or by simply filling the $N$
lowest-energy single-electron states in two dimensions.

The operator $T$ is a simple many-body excitation operator, which
in all standard coupled-cluster approximations is truncated at
a given (usually low)
$M$-particle-$M$-hole excitation level $M < N$, with $N$ being the number of electrons.  
If all excitations are included up to the 
$N$-particle-$N$-hole set of Slater determinants, one ends up with solving the full problem.
The general form of the truncated
cluster operator, defining a standard
single-reference coupled-cluster approximation characterized
by the chosen excitation level $M$, is
\begin{equation}
T=T_1+T_2+T_3+\dots +T_M,
\label{a5}
\end{equation}
where
\[
T_k =
\frac{1}{(k!)^2} \sum_{i_1,\ldots,i_k; a_1,\ldots,a_k} t_{i_1\ldots
i_k}^{a_1\ldots a_k}
\hat{a}^\dagger_{a_1}\ldots \hat{a}^\dagger_{a_k}
\hat{a}_{i_k}\ldots \hat{a}_{i_1} \ .
\]
Here and in the following, the indices $i, j, k,\ldots$ label occupied single-particle 
orbitals
while $a,b,c,\ldots$ label unoccupied orbitals.
The unknown amplitudes $t_i^a$, $t_{ij}^{ab}$ etc
in the last equation are determined from the solution of the coupled-cluster equations discussed below.
For a truncated $T$ operator, we will use the notation $T(M)$, where $M$ refers to highest possible
particle-hole excitations.

As an example, we list here the expressions for one-particle-one-hole, two-particle-two-hole, and
three-particle-three-hole  operators, labeled $T_1$, $T_2$, and $T_3$, respectively,
\begin{equation}
T_1 = \sum_{i<\varepsilon_f}\sum_{a>\varepsilon_f} t^a_i \hat{a}^{\dagger}_a \hat{a}_i 
\label{eq:t1}
\end{equation}
and 
\begin{equation}
T_2 = \frac{1}{4}\sum_{ij<\varepsilon_f}\sum_{ab > \varepsilon_f}t^{ab}_{ij}\hat{a}^{\dagger}_a \hat{a}^{\dagger}_b \hat{a}_j \hat{a}_i,
\label{eq:t2}
\end{equation}
and finally
\begin{equation}
T_3 = \frac{1}{36}\sum_{ijk<\varepsilon_f}\sum_{abc > \varepsilon_f}t^{abc}_{ijk}\hat{a}^{\dagger}_a \hat{a}^{\dagger}_b\hat{a}^{\dagger}_c \hat{a}_k\hat{a}_j \hat{a}_i.
\label{eq:t3}
\end{equation}
We will in this work limit ourselves to a single reference Slater determinant $\Phi_0$.

The cluster amplitudes $t_{i_1\ldots i_n}^{a_1\ldots a_n}$
are determined by solving a coupled system of nonlinear and energy-independent
algebraic equations of the form
\begin{equation}
\langle \Phi_{i_{1} \ldots i_{n}}^{a_{1} \ldots a_{n}} |
\bar{H}|\Phi_0\rangle = 0 , \;\;\;\; i_{1}< \cdots < i_{n}, \;\;
a_{1} < \cdots < a_{n},
\label{ccaeq}
\end{equation}
where $n=1,\ldots,M$. Here,
\begin{equation}
\bar{H} = e^{-T(M)} \hat{H} \, e^{T(M)}
= (\hat{H} \, e^{T(M)})_{C},
\label{hbara}
\end{equation}
is the similarity-transformed
Hamiltonian of the coupled-cluster theory truncated at $M$-particle-$M$-hole
excitations and the subscript $C$ denotes
the connected part of the corresponding
operator expression, 
and $|\Phi_{i_{1} \ldots i_{n}}^{a_{1} \ldots a_{n}} \rangle
\equiv a^{a_{1}} \cdots a^{a_{n}} a_{i_{n}} \cdots a_{i_{1}}
|\Phi \rangle$ are the $n$-particle-$n$-hole  or $n$-tuply
excited determinants relative to reference determinant $|\Phi_0\rangle$.
The Hamiltonians $\bar{H}$ and $\hat{H}$ are normal ordered.

If we limit ourselves to include only one-particle-one-hole and two-particle-two-hole excitations,
what is known as coupled cluster of singles and doubles (CCSD), the 
method corresponds to $M=2$, and the cluster operator $T^{(N)}$
is approximated by
\begin{equation}
T(M) =T(2) = T_1 + T_2 ,
\label{eq:t1t2}
\end{equation}
given by the operators of Eqs.~(\ref{eq:t1}) and (\ref{eq:t2}).

The standard CCSD equations for the singly and doubly
excited cluster amplitudes
$t_a^i$ and $t_{ab}^{ij}$, defining $T_1$ and $T_2$, respectively,
can be written as
\begin{equation}
\langle \Phi_{i}^{a} | \bar{H}{\rm (CCSD)}|\Phi_0\rangle = 0,
\label{ccsd1}
\end{equation}
and
\begin{equation}
\langle \Phi_{ij}^{ab} | \bar{H}{\rm (CCSD)}|\Phi\rangle = 0, \;\;\;
i < j, \; a < b ,
\label{ccsd2}
\end{equation}
where
\begin{equation}
\bar{H}{\rm (CCSD)} =\bar{H} = e^{-T(2)}\hat{H} e^{T(2)}=(\hat{H} e^{T(2)})_{C}
\label{hccsd}
\end{equation}
is the similarity-transformed Hamiltonian of the CCSD approach and the subscript $C$ stands for
connected diagrams only.

The system of coupled-cluster equations is obtained
in the following way. We first insert
the coupled-cluster wave function $|\Psi_{0} \rangle$
into the $N$-body Schr{\" o}dinger equation,
\begin{equation}
\hat{H} |\Psi_{0} \rangle = \Delta E_{0} |\Psi_{0} \rangle ,
\label{schreq}
\end{equation}
where
\[
\Delta E_{0} = E_{0} - \langle \Phi_0 | \hat{H} | \Phi_0 \rangle
\label{deltae}
\]
is the corresponding energy relative to the
reference energy $\langle \Phi_0 | \hat{H} | \Phi_0 \rangle$,
and premultiply both sides
on the left by $e^{-T^{(N)}}$ to obtain
the connected-cluster form of the Schr{\" o}dinger equation
\begin{equation}
\bar{H} |\Phi \rangle = \Delta E_{0} |\Phi \rangle ,
\label{rightcc}
\end{equation}
where
\begin{equation}
\bar{H} = e^{-T(2)} \hat{H} \, e^{T(2)}
= (H \, e^{T(2)})_{C}
\label{similaritycc}
\end{equation}
is the similarity-transformed Hamiltonian.

Next, we project Eq.~(\ref{rightcc}), in which
$T$ is replaced by its approximate form
$T(M)$, Eq.~(\ref{a5}), onto the excited determinants
$|\Phi_{i_{1} \ldots i_{n}}^{a_{1} \ldots a_{n}} \rangle$,
corresponding
to the $M$-particle-$M$-hole excitations included in $T_M$.
The excited determinants
$|\Phi_{i_{1} \ldots i_{n}}^{a_{1} \ldots a_{n}} \rangle$ are orthogonal
to the reference determinant $|\Phi_0\rangle$, so that
we end up with nonlinear and energy-independent
algebraic equations of the form of Eq.~(\ref{ccaeq}).

Once the system of equations, Eq. (\ref{ccaeq}),
is solved for $T_M$ or $t_{a_{1}\ldots a_{n}}^{i_{1} \ldots i_{n}}$
(or Eqs.~(\ref{ccsd1}) and (\ref{ccsd2})
are solved for $T_{1}$ and $T_{2}$ or $t_a^i$ and $t_{ab}^{ij}$),
the ground-state coupled-cluster energy is calculated using the equation
\begin{eqnarray}
E_{0} & = & \langle \Phi_0 | \hat{H} | \Phi_0 \rangle + \Delta E_{0}
\nonumber \\
& = &
\langle \Phi_0 | \hat{H} | \Phi_0 \rangle
+ \langle\Phi_0|\bar{H}|\Phi_0\rangle.
\end{eqnarray}

It can easily be shown that if $H$
contains only up to two-body interactions
and $2 \leq M \leq N$, we can write
\begin{eqnarray}
E_{0} = \langle \Phi_0 | \hat{H} | \Phi_0 \rangle +
\langle \Phi_0 | [ \hat{H}(T_{1} + T_{2} + \frac{1}{2} T_{1}^{2}) ]_{C} | \Phi_0 \rangle .
\label{egratwobody}
\end{eqnarray}
In other words, we only need
$T_{1}$ and $T_{2}$ clusters to calculate the ground-state energy
$E_{0}$ of the $N$-body ($N \geq 2$) system even if
we solve for other cluster components $T_{n}$ with $n > 2$. 
As long as the Hamiltonian contains up to two-body interactions,
the above energy expression is correct even in the exact case, when the
cluster operator $T$ is not truncated (see, for example, Refs.~\onlinecite{bartlett2007,helgaker2003} for proof).

The nonlinear character of the system of coupled-cluster
equations of the form of Eq.~(\ref{ccaeq}) does not mean that
the resulting equations contain very high powers of $T_M$.
For example, if the Hamiltonian $\hat{H}$ 
does not
contain higher--than--pairwise interactions,
the CCSD equations for the $T_{1}$ and $T_{2}$ clusters, or for the
amplitudes $t_a^i$ and $t_{ab}^{ij}$ that represent
these clusters, become
\begin{equation}
\langle \Phi_{i}^{a} |
[\hat{H}(1 + T_{1} + T_{2} + \frac{1}{2} T_{1}^{2} + T_{1} T_{2}
+ \frac{1}{6} T_{1}^{3} )]_{C}
|\Phi \rangle = 0,
\label{momccsd1}
\end{equation}
\begin{eqnarray}
\langle \Phi_{ij}^{ab} | && \!\!\!\!\!\!\!
[\hat{H}(1 + T_{1} + T_{2} + \frac{1}{2} T_{1}^{2} + T_{1} T_{2}
+ \frac{1}{6} T_{1}^{3}
\nonumber \\
&& + \frac{1}{2} T_{2}^{2} + \frac{1}{2} T_{1}^{2} T_{2}
+ \frac{1}{24} T_{1}^{4} )]_{C}
|\Phi \rangle = 0.
\label{momccsd2}
\end{eqnarray}

The explicitly connected form of the coupled-cluster equations,
such as Eqs.~(\ref{ccaeq}) or (\ref{momccsd1}) and (\ref{momccsd2}),
guarantees that the process of solving these equations
leads to connected terms in cluster components of $T$ and
connected terms in the energy $E_{0}$, independent of
the truncation scheme $M$ used to define $T_M$.
The absence of disconnected terms in $T_M$ and $E_{0}$
is essential to obtain the rigorously size-extensive results \cite{bartlett2007}.
It is easy to extend the above equations for the cluster amplitudes to include
triples excitations, leading to the so-called CCSDT\cite{ccsdt-n} hierachy of equations. 
Defining 
\[
  f= \sum_{pq}f_{pq}\{a^+_p a_q\}
\]
with $f_{pq}$ the Fock matrix elements  and 
\[
   W=\frac{1}{4}\sum_{pqrs}\langle pq||rs \rangle \{a^+_p a^+_q a_r a_s\}
\]
where $\langle pq||rs \rangle$ are anti-symmetrized two-body matrix elements, 
the extension to triples gives  the following equations for the amplitudes with 
one-particle-one-hole excitations
\begin{widetext}
\[
\langle \Phi_{i}^{a} |
[fT_1+ f(T_{2} + 1/2 T_{1}^{2}) +WT_1 +W(T_{2} + 1/2 T_{1}^{2})+ W(T_{1} T_{2}
+ 1/6 T_{1}^{3}+T_3 )]_{C}
|\Phi \rangle = 0,
\]
\end{widetext}
and with two-particle-two-hole excitations
\begin{widetext}
\[
\langle \Phi_{ij}^{ab} |
[fT_1+f(T_3+T_2T_1)+W+WT_1+W(T_{2} + 1/2 T_{1}^{2})
+W(T_{1} T_{2}
+ 1/6 T_{1}^{3}+T_3 )
\]
\[
+W(T_{1} T_{3}+ 1/2 T_{2}^{2}+ 1/2 T_2T_{1}^{2}
+ 1/24 T_{1}^{4})]_{C}
|\Phi \rangle = 0.
\]
\end{widetext}
and with three-particle-three-hole excitations we end up with
\begin{widetext}
\[
\langle \Phi_{ijk}^{abc} |
[fT_3+f(T_3T_1+1/2T_2^2)+ WT_2+W(T_{3} + T_{1}T_2)
+W(1/2T_{2}+ T_3T_1 1/2 T_{1}^{2}+T_1 )
\]
\[
+W(T_{2} T_{3}+ 1/2 T_{2}^{2}T_1+ 1/2 T_3T_{1}^{2}
+ 1/6 T_2T_{1}^{3})]_{C}
|\Phi \rangle = 0.
\]
\end{widetext}
Different approximations to the solution of the triples equations yield different CCSDT approximations.
The CCSD method scales (in terms of the most computationally expensive contributon) as $n_o^2n_u^4$, where $n_0$ represents the number of occupied orbitals and $n_u$ the number of unoccupied single-particle states. The  full CCSDT scales as  $n_o^3n_u^5$.

Coupled-cluster theory with inclusion of full triples CCSDT is
usually considered to be too computationally expensive in most many-body 
systems of considerable size. Therefore triples corrections are 
usually taken into account perturbatively using the non-iterative CCSD(T)
approach described in Ref.~\onlinecite{Deegan94}. Recently, a more
sophisticated way of including the full triples known as
the $\Lambda$-CCSD(T) approach, has been developed by 
Taube \emph{et al.} \cite{Kucharski98,Taube08,taube2}. In the
$\Lambda$-CCSD(T) approach, the left-eigenvector solution of the CCSD
similarity-transformed Hamiltonian is utilized in the calculation of a
non-iterative triples correction to the coupled-cluster ground-state
energy. The left eigenvalue problem is given by
\begin{equation}
\label{left}
\langle\Phi_0| \Lambda \bar{H} = E\langle\Phi_0| \Lambda \ ,
\end{equation}
were ${\Lambda}$ denotes the de-excitation cluster operator
\begin{equation}
{\Lambda} = 1 + {\Lambda}_1 +{\Lambda}_2 \ ,
\end{equation}
with
\begin{eqnarray}
{\Lambda}_1 &=&
\sum_{i,a} \lambda^{i}_{a}
a_{a}
a^\dagger_{i} \ , \\
{\Lambda}_2 &=&
\frac{1}{4} \sum_{i,j,a,b} \lambda^{ij}_{ab}
a_{b} a_{a}
a^\dagger_{i} a^\dagger_{j} \ .
\end{eqnarray}
The unknowns, $\lambda^i_a$ and $\lambda^{ij}_{ab}$, result from the ground-state
solution of the left eigenvalue problem (\ref{left}). Using a single-particle basis
that diagonalizes the fock matrix $f$ witin the $hole$-$hole$ and $
particle$-$particle$ blocks simultaneously, and utilizing the $\lambda^i_a$
and $\lambda^{ij}_{ab}$ de-excitation amplitudes together with the cluster amplitudes, $t_i^a$ and
$t_{ij}^{ab}$, we get the non-iterative $\Lambda$-CCSD(T) 
energy correction to the coupled-cluster correlation energy 
(see Ref.~\onlinecite{Taube08} for more details), 
\begin{eqnarray}
\Delta E_3&=& {1\over (3!)^2} \sum_{i j k a b c} \langle\Phi_0|
{\Lambda}({f}_{hp} +{W})_N|\Phi^{abc}_{ijk}\rangle\nonumber\\
&\times& {1\over
\gamma_{ijk}^{abc}}\langle\Phi_{ijk}^{abc}|({W}_N{T}_2)_C|\Phi_0\rangle
\ .  
\end{eqnarray}
Here, ${f}_{hp}$ denotes the part of the normal-ordered one-body
Hamiltonian that annihilates particles and creates holes, while 
\begin{equation}
\gamma_{ijk}^{abc} \equiv f_{ii}+f_{jj}+f_{kk}-f_{aa}-f_{bb}-f_{cc}
\end{equation}
is expressed in terms of the diagonal matrix elements of the
normal-ordered one-body Hamiltonian ${f}$. In the case of
Hartree-Fock orbitals, the one-body part of the Hamiltonian is
diagonal and ${f}_{hp}$ vanishes. The state $|\Phi_{ijk}^{abc}\rangle$ is a three-particle-three-hole 
excitation of the reference state.
For a further discussion of various approximations to the triples correlations, see, for example,
the recent review by Bartlett and Musia{\l} \cite{bartlett2007}.

In this work we will focus on the CCSD, the CCSD(T), and the
$\Lambda$-CCSD(T) approaches, using either a renormalized or an
unrenormalized interaction. In order to avoid an iterative solution of
the CCSD(T) and $\Lambda$-CCSD(T) equations, we start from a self-consistent
Hartree-Fock basis such that the Fock matrix $f$ is diagonal.
Using such a basis, the computational cost of the CCSD(T) and
$\Lambda$-CCSD(T) energy corrections is $n_o^3n_u^4$ number of cycles, done only once.   
It is also important to keep in mind, in particular when linking our coupled-cluster
theory with Monte Carlo approaches, that a wavefunction-based method 
like coupled-cluster theory is defined within a specific subset of 
the full Hilbert space. In our case, the Hilbert space will be defined 
by all possible many-body wave functions which can be constructed
within a certain number of the lowest-lying single-particle states.

\subsection{Diffusion  Monte Carlo}\label{subsec:dmc}

The diffusion Monte Carlo method seeks the solution of the equation:
\begin{equation}
\partial_{\tau}|\Psi({\bf R},\tau)\rangle=[\hat H-E_{0}]|\Psi(R,\tau)\rangle
\label{imSeq}
\end{equation}
where ${\bf R}$ collectively indicates the degrees of freedom of the system (the 3N
electron coordinates, in this case).
By expanding the state $|\Psi({\bf R},\tau)\rangle$ on the basis of eigenstates $|\phi_{n}\rangle$
of $\hat H$, a formal solution of Eq. (\ref{imSeq}) is given by
\begin{equation}
\begin{array}{rcl}
|\Psi({\bf R},\tau)\rangle&=&e^{-(\hat H-E_{0})\tau}|\Psi({\bf R},0)\rangle\\
&=&\sum_{n}e^{-(\hat H-E_{0})\tau}|\phi_{n}\rangle
\langle\phi_{n}|\Psi({\bf R},0)\rangle\\
&=&\sum_{n}e^{-(\hat E_{n}-E_{0})\tau}|\phi_{n}\rangle\langle\phi_{n}|\Psi({\bf R},0)\rangle
\end{array}
\end{equation}
from which it is evident that for $\tau \rightarrow \infty$ the only surviving component
is the ground state of $\hat H$. Eq. (\ref{imSeq}) can be numerically solved by expanding
the state to be evolved in eigenstates $|{\bf R}_{i}\rangle$ of the position operator (called ``walkers''), so
that the evolution reads:
\begin{equation}
\sum_{i}\langle {\bf R}_{i}|\Psi({\bf R},\tau)\rangle = \sum_{i}\langle {\bf R}|e^{-(\hat H-E_{0})\tau}|{\bf R}'_{i}\rangle\langle {\bf R}'_{i}|\Psi({\bf R}',0)\rangle.
\end{equation}
Formally, in terms the Green's function solution of Eq. (\ref{imSeq}), the solution can be written as:
\begin{equation}
\Psi({\bf R},\tau)=\int  G({\bf R}', {\bf R},\tau)\Psi({\bf R}',0)\;d{\bf R}' .
\end{equation}
The  Green's function $G({\bf R}', {\bf R},\tau)=\langle {\bf R}|\exp[-(\hat H-E_{0})]| {\bf R} \rangle$ is in general unknown. However, in the limit $\Delta\tau\rightarrow 0$ it can be written in the following form:
\begin{equation}
G({\bf R}',{\bf R},\tau)\simeq\sqrt{\left(\frac{ m_{e}m^{*}}{2\pi\hbar^{2}\Delta\tau}\right)^{d}}
e^{\frac{({\bf R}-{\bf R}')^{2}}{2\hbar^{2}/m_{e}m_{*}\Delta\tau}}e^{-[V({\bf R})-E_{0}]\Delta\tau},
\end{equation}
that is as the product of the free particle Green's function,
having the effect of displacing the $d$-dimensional walkers, and a factor containing the potential, which is interpreted as a weight
for the estimators computed at the walker position, and a probability for the walker itself to generate one or
more copies of itself in the next generation. Due to the divergence of the potential at the origin,
it is necessary to modify the algorithm, introducing the so-called ``importance sampling''. In 
practice, the sampled distribution is modified by multiplying by an approximate solution of the Schr\"odinger
equation $\Psi_{T}({\bf R})$, which is usually determined by a variational Monte Carlo calculation:
\begin{equation}
\Psi_{T}({\bf R})\Psi({\bf R},\tau)=\int  G({\bf R}', {\bf R},\tau)\frac{\Psi_{T}({\bf R})}{\Psi_{T}({\bf R}')}\Psi_{T}({\bf R}')\Psi({\bf R}',0)\;d{\bf R}'.
\end{equation}
A final  important observation is the fact that the procedure described above is well defined only in the case
of a totally symmetric ground state. For a many-Fermion system it would be necessary, in principle, to project
on an excited state of the Hamiltonian, which leads to a severe instability of the variance on the energy
estimation. This problem is usually treated by artificially imposing, as an artificial boundary condition, that
the solution vanishes on the nodes of the trial function $\Psi_{T}$ (fixed-node approximation). Many other technical
details enter the real calculation. A thorough description of the DMC algorithm, as implemented for
the calculations of this paper, can be found in Ref.~\onlinecite{Umr93}.

The  fixed-node DMC calculations depend on the
quality of the trial wavefunction $\Psi_{T}({\bf R})$, which is usually built starting from a parametrized ansatz. The 
values of the parameters are computed by minimizing the expectation value of the Hamiltonian
on $\Psi_{T}({\bf R})$.
The trial wavefunctions we use have the form\cite{pederiva2001}:
\begin{equation}
\Psi({\bf R})_{L,S}=
\exp[\phi({ R})]\sum_{i=1}^{N_{\rm conf}} \alpha_i\Xi_i^{L,S}({\bf R})~~,
\label{eq:eq2}
\end{equation}
where the $\alpha_i$ are variational parameters. Because in this paper we are considering only
closed--shell dots which have $L=0$ and $S=0$, the sum in Eq. (\ref{eq:eq2}) reduces to a single term:
\begin{equation}
\Xi^{L=0,S=0}= D^\uparrow D^\downarrow~~,
\label{eq:eq3}
\end{equation}
where the $D^\chi$ are Slater determinants of spin-up and spin-down electrons,
using orbitals from a local density approximation  calculation with the same confining potential
and the same number of electrons. 
The function $\exp[\phi]$ in Eq. (\ref{eq:eq2}) is a generalized Jastrow
factor of the form:
\begin{eqnarray}
\begin{array}{l}
\phi(R)=\displaystyle\sum_{i=1}^{N}\left[\sum_{k=1}^{6}\gamma_k J_0\left(
\frac{k\pi r_i}{R_c}\right)\right] +\\
\displaystyle\sum_{i<j}^N\frac{1}{2}\left(\frac{a_{ij} r_{ij}}{1+b(r_i)r_{ij}}+
\frac{a_{ij} r_{ij}}{1+b(r_j)r_{ij}}\right)~,
\end{array}
\end{eqnarray}
where
\begin{equation}
b(r)=b_0^{ij}+b_1^{ij}\tan^{-1}[(r-R_c)^2/2R_c\Delta].
\end{equation}
It explicitly includes one- and two-body correlations and effective
many-body correlations via the spacial dependence of $b(r)$.
The quantity $R_c$ represents an ``effective'' radius of the dot, which
is optimized in the variational procedure.
The $b_0$ and $b_1$ parameters depend only on the {\it relative} spin
configuration of the pair $ij$.
The parameters $a_{ij}$ are fixed in order to satisfy the
cusp conditions, that is, the condition of finiteness of
the local energy $\hat H\Psi/\Psi$ for $r_{ij}\rightarrow 0$. For a two
dimensional system, $a_{ij}=1$ if the electron pair $ij$ has antiparallel
spin, and $a_{ij}=1/3$ otherwise.
The dependence of $a_{ij}$ on the relative spin orientation of the electron
pair introduces spin-contamination into the wavefunction.
However, the magnitude of the spin contamination and its effect on the
energy has been shown to be totally negligible in the case of well-optimized
atomic wavefunctions~\cite{Chi98} and we expect that to be true here as well.

Also the coefficients $\gamma_k$ in the one--body term,
the coefficients $\Delta$, $b_0$, and $b_1$ in the two--body term, and
the coefficients $\alpha_i$ multiplying the configuration state functions
are optimized by minimizing the variance of the local energy~\cite{UWW88}.

\section{Results}\label{sec:results}
We start our discussion with the results for the two-electron system, 
since these can, for certain values of the oscillator frequency, 
be compared with the exact results of Taut \cite{taut1994}. These 
results are presented in the next subsection using both a renormalized 
two-body Coulomb interaction and the `bare' Coulomb interaction.  
Thereafter we present coupled-cluster results with singles and doubles 
excitations for systems with $N=6$ and $N=12$ electrons with the bare Coulomb interaction.
The slow convergence as a function of the number of oscillator shells with the bare interaction 
serves to motivate the introduction of an effective Coulomb
interaction. In the main result section, we present CCSD, CCSD(T), and
$\Lambda$-CCSD(T) results for $N=6,12$ and  $N=20$ electrons using an
effective two-body Coulomb interaction and compare with 
diffusion Monte Carlo (DMC) calculations for the same systems.  

\subsection{Results for two electrons}
In this subsection we limit our attention to the two-electron system
and compare our DMC results with
coupled-cluster calculations with CCSD correlations only.
The results presented here serve to demonstrate the reliability of 
using an effective Coulomb interaction.

The CCSD approach gives the exact eigenvalues for the two-particle
system. We have employed a standard harmonic oscillator basis using 
the frequencies $\omega=0.5$ and $\omega=1.0$ a.u. Our results are 
listed in Table \ref{table:tab1}. The variable $R$ represents the 
number of oscillator shells in which the effective interaction case 
represents the model space for which the effective
Coulomb interaction is defined. The calculations labeled CCSD-$V$ 
represent the results obtained with the unrenormalized or bare Coulomb 
interaction, while the shorthand CCSD-$V_{\mathrm{eff}}$ stands for the results 
obtained with an effective interaction. Since the latter, irrespective 
of size of model space (number of lowest-lying oscillator shells in our case) always 
gives the exact lowest-lying eigenvalues by construction 
(a similarity transformation preserves always the eigenvalues), 
these results are unchanged as a function of the number of oscillator
shells $R$. For the two-body problem, coupled-cluster theory at the level
of singles and doubles excitations yields the same as exact
diagonalization in the same two-particle space. In our case, the
number of two-body configurations is given by all allowed
configurations that can be constructed by placing two particles in the 
single-particle orbits defined by the given number of oscillator shells $R$. 
\begin{table}[hbtp]
  \begin{tabular}{|c|c|r|r|r|}\hline
    \multicolumn{1}{|c}{$\omega$} & \multicolumn{1}{|c}{$R$} & \multicolumn{1}{|c}{ CCSD-$V$ }& \multicolumn{1}{|c}{ CCSD-$V_{\mathrm{eff}}$ }& \multicolumn{1}{|c|} {DMC} \\
    \hline
                0.5       & 2  & 1.786914 & 1.659772 & \\
                       & 4  & 1.673874 & 1.659772 & \\
                       & 6  & 1.667259 & 1.659772 & \\
                       & 8  & 1.664808 & 1.659772 & \\
                       & 10 & 1.660211 & 1.659772 & \\
                       & 12 & 1.660091 & 1.659772 &\\
                       & 14 & 1.660018 & 1.659772 & \\
                       & 16 & 1.659970 & 1.659772 & 1.65975(2)\\\hline
                1.0       & 2  & 3.152329 & 3.000000 & \\
                       & 4  & 3.025232 & 3.000000 & \\
                       & 6  & 3.013627 & 3.000000 & \\
                       & 8  & 3.009237 & 3.000000 & \\
                       & 10 & 3.000895  & 3.000000 & \\
                       & 12 & 3.000654 & 3.000000 & \\
                       & 14 & 3.000505 & 3.000000 & \\
                       & 16 & 3.000406 & 3.000000 & 3.00000(3)\\ \hline

  \end{tabular}
  \caption{Ground-state energies for two electrons in a circular quantum dot 
within the CCSD approach with (CCSD-$V_{\mathrm{eff}}$) and without (CCSD-$V$)
an effective Coulomb interaction. The diffusion Monte Carlo (DMC) results are also included. 
For $\omega=1$ Taut's exact result from Ref.~\onlinecite{taut1994} is
3 a.u. All energies are in atomic units. There are no triples corrections for the two-body problem.
The variable $R$ represents the number of oscillator shells.}
  \label{table:tab1}
\end{table}
For $\omega=1.0$ a.u.,Taut's exact result from Ref.~\onlinecite{taut1994} is reproduced. 
The non-interacting
part of the Hamiltonian gives a contribution of $2$ a.u.~to the ground-state energy while the 
two-particle interaction results in $1$ a.u.

We notice also that the DMC results agree perfectly (within six leading digits) with our
CCSD-$V_{\mathrm{eff}}$ calculations. The standard error in the DMC calculations is given in parentheses. 

If we, on the other hand, use the bare Coulomb interaction, we see that the convergence of the 
CCSD-$V$ results as a function of $R$ is much slower and in line with the analysis of Ref.~\onlinecite{kvaal2009}
and our discussion in 
subsection \ref{subsec:effint}. One needs at least some 16-20 major oscillator shells (between 
$272$ and $420$ single-particle states) in order to get a result within three to four leading digits
close to the exact answer. The slow convergence of the bare
interaction for the two-electron problem may be even more prevalent 
in a many-body system, in particular for small values of $\omega$, 
where correlations are expected to be more important. 
With more particles, we may expect even worse convergence.  
In Table \ref{table:barecoulomb} we present for the case of
$\omega=1.0$ a.u.~CCSD results for  $N=6$ and $N=12$ electrons. 
The bare Coulomb interaction in an oscillator basis is used.  
The diffusion Monte Carlo results are for $N=6$ $20.1597(2)$ a.u.~and
for $N=12$ $65.700(1)$ a.u. Using the bare interaction thus results in 
a slow convergence, as will be demonstrated in the next subsection. 
The result of $20.1742$ a.u.~obtained with an effective Coulomb at the CCSD
level for $N=6$ and $R=10$ is much closer to the DMC result, as can be seen from Table
\ref{table:tab2}.
\begin{table}[hbtp]
  \begin{tabular}{|c|r|r|}\hline
\multicolumn{1}{|c}{$R$} & \multicolumn{1}{|c}{ $N=6$ }& \multicolumn{1}{|c|}{$N=12$ } \\
    \hline
                      2  & 22.219813 &  \\
                      3  & 21.419889 &  73.765549 \\ 
                      4  & 20.421325 & 70.297531  \\
                      5  & 20.319716 & 66.989912 \\ 
                      6  & 20.260893 & 66.452006 \\
                      7  & 20.236760 & 65.971686 \\
                      8  & 20.221750 & 65.889324 \\
                      9  & 20.211590 & 65.838932  \\
                      10 & 20.204345 & 65.806539 \\\hline

  \end{tabular}
  \caption{Ground-state energies for $N=6$ and $N=12$ electrons in a circular quantum dot
within the CCSD approach using the bare Coulomb interaction.
All energies are in atomic units. There are no triples corrections. Results are presented for an oscillator frequency $\omega = 1.0$ a.u. The variable $R$ represents the number of oscillator shells. For $N=12$ the first three shells are filled and there are no results for two shells only.}
  \label{table:barecoulomb}
\end{table}
These results serve the aim of motivating the introduction of an effective two-particle interaction.
In the next subsection, we will make further comparisons between our results with and without an effective interaction. In particular, we will try to extract convergence criteria for both approaches and link
our numerical results with the predictions made by Kvaal in Eq.~(\ref{eq:simenproof}).

\subsection{Results with an effective Coulomb interaction}
We present here our final and most optimal results for $N=6$, $N=12$, 
and $N=20$ electrons using the CCSD, the CCSD(T), and the 
$\Lambda$-CCSD(T) approaches. We list the CCSD(T) triples results as well. This method has for a long time
been considered as the calculational  'gold standard' in quantum
chemistry due to its low computational cost and accuracy. We
emphasize, however, that the $\Lambda$-CCSD(T) approach is an 
improvement of the standard CCSD(T) approach, and should therefore 
be considered as our best and most accurate coupled-cluster
calculation in this work. In all calculations we employ  
an effective Coulomb interaction and a self-consistent Hartree-Fock
basis for different values of the oscillator frequency $\omega$ and the 
model space $R$. The results are compared with diffusion Monte Carlo 
calculations (DMC)\cite{Note_DMC}.  In addition to 
the values of $\omega=1.0$ and $\omega=0.5$, which serve more as a
reference for earlier calculations, we present results for
$\omega=0.28$ a.u., which corresponds to 3.32 eV,  a frequency which should approximate the
experimental situation in Ref.~\onlinecite{Tar96}.
 The role of correlations is also more important for smaller values of
$\omega$, allowing us therefore to test the reliability of
our single-reference CCSD and $\Lambda$-CCSD(T) calculations. As the
system becomes more and more correlated, contributions from clusters beyond the $T(3)$ (beyond three-particle-three-hole correlations) 
level might become non-neglible. For values of $\omega > 1$, 
the single-particle part of the Hamiltonian dominates and correlations 
play a less prominent role.

Our results for $N=6$, $N=12$, and $N=20$ electrons are displayed 
in Tables \ref{table:tab2}, \ref{table:tab3}, and \ref{table:tab4} 
respectively. We present also the mean-field energies 
(that is, the Hartree-Fock ground-state energies). These are labeled as
$E_0$ in the Tables. For all values of $\omega$ with $R=20$ major
oscillator shells, our best coupled-cluster results, the 
$\Lambda$-CCSD(T) calculations, are very close to the diffusion 
Monte Carlo calculations. Even for $10$ major shells, the results 
are close to the DMC calculations, suggesting thereby that the usage 
of an effective interaction provides a better starting point for 
many-body calculations. The convergence of the coupled-cluster
calculation in terms of the number of major oscillator shells is 
also better than the results shown in Table \ref{table:barecoulomb}
with the bare Coulomb interaction. This discussion will be further elaborated at the end of this section.
\begin{table}[hbtp]
  \begin{tabular}{|c|c|r|r|r|r|r|}\hline
    \multicolumn{1}{|c}{$\omega$} & \multicolumn{1}{|c}{$R$}& \multicolumn{1}{|c}{$E_0$ } & \multicolumn{1}{|c}{ CCSD }& \multicolumn{1}{|c}{ CCSD(T) }& \multicolumn{1}{|c}{ $\Lambda$-CCSD(T) }&
    \multicolumn{1}{|c|} {DMC} \\
    \hline
0.28&  10&       7.9504&       7.6241&       7.6032&       7.6064& \\
&  12&       7.9632&       7.6245&       7.6023&       7.6057& \\
&  14&       7.9720&       7.6247&       7.6016&       7.6052& \\
&  16&       7.9785&       7.6249&       7.6012&       7.6048& \\
&  18&       7.9834&       7.6251&       7.6008&       7.6046& \\
&  20&       7.9872&       7.6252&       7.6006&       7.6044& 7.6001(1)\\\hline
0.5&  10&      12.1927&      11.8057&      11.7871&      11.7892& \\
&  12&      12.2073&      11.8055&      11.7858&      11.7880& \\
&  14&      12.2173&      11.8055&      11.7850&      11.7873& \\
&  16&      12.2246&      11.8055&      11.7845&      11.7868& \\
&  18&      12.2302&      11.8055&      11.7841&      11.7864& \\
&  20&      12.2346&      11.8055&      11.7837&      11.7862& 11.7888(2)\\ \hline
1.0&  10&      20.6295&      20.1766&      20.1623&      20.1633& \\
&  12&      20.6461&      20.1753&      20.1602&      20.1612& \\
&  14&      20.6576&      20.1746&      20.1589&      20.1600& \\
&  16&      20.6659&      20.1742&      20.1580&      20.1592& \\
&  18&      20.6723&      20.1739&      20.1574&      20.1586& \\
&  20&      20.6773&      20.1737&      20.1570&      20.1582& 20.1597(2)\\ \hline
  \end{tabular}
  \caption{Ground-state energies for $N=6$ electrons in a circular quantum dot
within various coupled-custer approximations utilizing 
an effective Coulomb interaction and the diffusion Monte Carlo (DMC) approach. 
The coupled-cluster results have been obtained with an effective
two-body interaction using a self-consistent Hartree-Fock basis and
the CCSD, the CCSD(T), and the $\Lambda$-CCSD(T) approaches discussed 
in the text. $E_0$ is the Hartree-Fock energy while $R$ stands for the
number of major oscillator shells. All energies are in atomic units.}
  \label{table:tab2}
\end{table}

In $R=20$ major shells the $\Lambda-$CCSD(T) results are very close to the DMC
results. As an example, consider the $\omega=1$ results for $N=6$ in 
Table \ref{table:tab2}. The CCSD result is 20.1737 a.u., while the 
$\Lambda-$CCSD(T) number is 20.1582 a.u. The corresponding DMC 
energy is 20.1597(2) and very close to our $\Lambda-$CCSD(T)
result. With $R=20$ shells our coupled-cluster calculations are almost converged at the level of
the fifth or sixth  number after the decimal point. At the end of this section we discuss
the convergence properties of the various coupled-cluster approaches as functions of
the number of oscillator shells $R$. 

In Tables \ref{table:tab2}, \ref{table:tab3}, and \ref{table:tab4} we
see that the CCSD(T) results are in most cases overshooting the 
diffusion Monte Carlo results. From numerous coupled-cluster studies 
in quantum chemistry, it has been found that CCSD(T) tends to 
overestimate the role of triples and thereby often
overshoots the exact energy. The $\Lambda$-CCSD(T) approach has, on the 
other hand, been found to give highly accurate correlation energies, and 
even in some cases performing better than the full CCSDT approach
(see Refs. \cite{Taube08, taube2}). This is also consistent with our
findings for the CCSD(T) and $\Lambda$-CCSD(T) correlation energies in 
quantum dots. 

Let us briefly discuss the error in our coupled-cluster calculations.  
There are two sources of error, the first coming from the finite size
of the single-particle basis, and the other from truncation of the
cluster amplitude $T$ at the $T(3)$ excitation level
(three-particle-three-hole exciations). We are presently not able 
to provide a  mathematical error estimate on truncations in terms 
of the number of particle-hole excitation operators in the cluster 
operator $T$. However, several studies from quantum chemistry (see Ref. 
\cite{bartlett2007} and references therein) and in nuclear physics
\cite{hagen2009, hagen2010b} have shown that the CCSD approach gives about 90$\%$ of the
correlation energy while CCSDT gives about 99$\%$ of the full
correlation energy. Assuming that the DMC results are to be considered as 
exact results, we can calculate the percentage of correlation energy
our CCSD and $\Lambda$-CCSD(T) calculations give for different numbers of
electrons $N$ and values $\omega$ of the confining harmonic oscillator potential.  
In Table \ref{table:tab5} we list the amount (in percentage) of 
correlation energy obtained at the CCSD and $\Lambda$-CCSD(T)
level; the coupled-cluster calculations were done in a 
model space of $R=20$ major oscillator shells. 

As we see from Table~\ref{table:tab5}, the CCSD approximation gives 90$\%$,
or more, of the full correlation energy, while the $\Lambda$-CCSD(T) 
approximation is at the level of 99-100$ \%$ of the full correlation
energy for $R=20$. The CCSD approximation is clearly performing better for larger 
values $\omega$ of the confining potential, but this is expected 
since the system becomes less and less correlated for larger 
values of $\omega$. This shows that our coupled-cluster calculations 
of circular quantum dots are within or even better than the accuracy 
seen in different applications in both quantum chemistry and nuclear
physics. 
    
As previously discussed, DMC results reported in this paper are still affected
by the fixed-node approximation. The extent of the error only depends on the
nodal surface of the wavefunction. Because we use a single product of 
Slater determinants, given the circular symmetry of the dots considered,
the nodes depend only on the set of single-particle functions used. 
Previous tests performed changing the set of single-particle orbitals show that
differences are of the order of one millihartrees or less\cite{pederiva2001}.
The optimization of the Jastrow factor only influences the variance of the energy,
which is typically of the order of 0.5\% of the total energy. 
Therefore, for circular quantum dots, we can conclude, assuming that 
the DMC calculations are as close as possible to the exact energies, 
that with an effective two-body interaction, a finite basis set of
$R=20$ major oscillator shells, and at most three-particle-three-hole 
correlations in the cluster amplitude, the remaining many-body effects 
are almost negligible as we are within 99-100\% of the full correlation energy.    

\begin{table}[hbtp]
  \begin{tabular}{|c|c|r|r|r|r|r|}\hline
    \multicolumn{1}{|c}{$\omega$} & \multicolumn{1}{|c}{$R$}& \multicolumn{1}{|c}{$E_0$ } & \multicolumn{1}{|c}{ CCSD}& \multicolumn{1}{|c}{ CCSD(T) }& \multicolumn{1}{|c}{ $\Lambda$-CCSD(T) }&
    \multicolumn{1}{|c|} {DMC} \\
    \hline
0.28&  10&      26.3556&      25.7069&      25.6445&      25.6540& \\
&  12&      26.3950&      25.7066&      25.6388&      25.6491& \\
&  14&      26.4221&      25.7074&      25.6363&      25.6470& \\
&  16&      26.4410&      25.7081&      25.6346&      25.6456& \\
&  18&      26.4551&      25.7085&      25.6334&      25.6446& \\
&  20&      26.4659&      25.7089&      25.6324&      25.6439& 25.6356(1)\\ \hline
0.5&  10&      39.9948&      39.2218&      39.1659&      39.1721& \\
&  12&      40.0409&      39.2203&      39.1599&      39.1667& \\
&  14&      40.0709&      39.2197&      39.1565&      39.1635& \\
&  16&      40.0922&      39.2195&      39.1543&      39.1615& \\
&  18&      40.1080&      39.2194&      39.1527&      39.1601& \\
&  20&      40.1202&      39.2194&      39.1516&      39.1591& 39.159(1)\\ \hline
1.0&  10&      66.6596&      65.7552&      65.7118&      65.7149& \\
&  12&      66.7106&      65.7484&      65.7017&      65.7051& \\
&  14&      66.7445&      65.7449&      65.6961&      65.6996& \\
&  16&      66.7686&      65.7430&      65.6926&      65.6963& \\
&  18&      66.7867&      65.7417&      65.6903&      65.6941& \\
&  20&      66.8006&      65.7409&      65.6886&      65.6924& 65.700(1)\\  \hline
  \end{tabular}
  \caption{Same caption as in Table~\ref{table:tab2} except the
    results are for $N=12$ electrons.}
  \label{table:tab3}
\end{table}

\begin{table}[hbtp]
  \begin{tabular}{|c|c|r|r|r|r|r|} \hline
    \multicolumn{1}{|c}{$\omega$} & \multicolumn{1}{|c}{$R$}& \multicolumn{1}{|c}{$E_0$ } & \multicolumn{1}{|c}{ CCSD}& \multicolumn{1}{|c}{ CCSD(T)}& \multicolumn{1}{|c}{ $\Lambda$-CCSD(T)}&
    \multicolumn{1}{|c|} {DMC} \\
    \hline
0.28&  10&      63.2588&      62.2851&      62.1802&      62.1946& \\
&  12&      63.2016&      62.0772&      61.9503&      61.9692& \\
&  14&      63.2557&      62.0634&      61.9265&      61.9466& \\
&  16&      63.3032&      62.0646&      61.9214&      61.9423& \\
&  18&      63.3369&      62.0656&      61.9181&      61.9395& \\
&  20&      63.3621&      62.0664&      61.9156&      61.9375& 61.922(2)\\\hline
0.5&  10&      95.2872&      94.0870&      93.9864&      93.9971& \\
&  12&      95.3407&      93.9963&      93.8818&      93.8944& \\
&  14&      95.4164&      93.9921&      93.8700&      93.8833& \\
&  16&      95.4676&      93.9904&      93.8632&      93.8771& \\
&  18&      95.5043&      93.9895&      93.8588&      93.8730& \\
&  20&      95.5320&      93.9891&      93.8558&      93.8702& 93.867(3)\\\hline
1.0&  10&     157.4356&     156.0128&     155.9324&     155.9381& \\
&  12&     157.5613&     155.9868&     155.8978&     155.9042& \\
&  14&     157.6437&     155.9740&     155.8795&     155.8863& \\
&  16&     157.7002&     155.9669&     155.8687&     155.8758& \\
&  18&     157.7413&     155.9627&     155.8618&     155.8690& \\
&  20&     157.7725&     155.9601&     155.8571&     155.8646& 155.868(6) \\\hline
  \end{tabular}
  \caption{Same caption as in Table~\ref{table:tab2} except the
    results are for $N=20$ electrons.}
  \label{table:tab4}
\end{table}

\begin{table}[hbtp]
  \begin{tabular}{|c|c|r|r|r|r|r|}\hline
    \multicolumn{1}{|c}{} & \multicolumn{2}{|c}{$\omega =0.28$}&
    \multicolumn{2}{|c}{$\omega =0.5$ } & \multicolumn{2}{|c|}{
      $\omega =1.0$ } \\
    \hline
    \multicolumn{1}{|c}{$N$} & \multicolumn{1}{|c}{$\Delta E_2$}&
    \multicolumn{1}{|c}{$\Delta E_3$ } & \multicolumn{1}{|c}{
      $\Delta E_2$ }& \multicolumn{1}{|c}{ $\Delta E_3$ }& 
    \multicolumn{1}{|c}{ $\Delta E_2$}&
    \multicolumn{1}{|c|} {$\Delta E_3$} \\
    \hline
6  & 94$\%$ & 99$\%$ & 96$\%$ & 100 $\%$ & 97$\%$ & 100$\%$  \\
12 & 91$\%$ & 99$\%$ & 94$\%$ & 100 $\%$ & 96$\%$ & 100$\%$  \\
20 & 90$\%$ & 99$\%$ & 93$\%$ & 100$\%$ & 95$\%$ &   100$\%$  \\\hline
  \end{tabular}
  \caption{Percentage of correlation energy at the CCSD level ($\Delta
    E_2$) and at the $\Lambda$-CCSD(T) level ($\Delta E_3$), for
    different numbers of electrons $N$ and values of the confining
    harmonic potential $\omega$. All numbers are for $R=20$.}
  \label{table:tab5}
\end{table}

In order to study the role of correlations as a function of the
oscillator frequency $\omega$ and the number of electrons, 
we define the relative energy
\begin{equation}
\epsilon = \left|\frac{E_{\mathrm{DMC}}-\langle \hat{H}_0\rangle}{E_{\mathrm{DMC}}}\right|,
\label{eq:dmch0}
\end{equation}
where $\langle \hat{H}_0\rangle$ is the expectation value of the 
one-body operator, the so-called unperturbed part of the
Hamiltonian. For $N=6$ this corresponds to an expectation value  
$\langle \hat{H}_0\rangle=10\omega$ for the one-body part of the 
Hamiltonian, while for $N=12$ and $N=20$ the corresponding numbers 
are $\langle \hat{H}_0\rangle=28\omega$ and $\langle \hat{H}_0\rangle=60\omega$, respectively.
Assuming that the diffusion Monte Carlo results are as close as 
possible to the true eigenvalues, the quantity $\epsilon$ measures 
the role of the two-body interaction and correlations caused by 
this part of the Hamiltonian as functions of $\omega$ and $N$, 
the number of electrons. The results for $\epsilon$ are shown in 
Fig.~\ref{fig:fig1}.  Results for $N=2$ are also included.

We see from this figure that the effect of the two-body interaction 
becomes increasingly important as we increase the number of
particles. Moreover, the interaction is more important for the smaller 
values of the oscillator frequency $\omega$. This is expected since
the contribution from the one-body operator is reduced due to smaller 
values of $\omega$. Including more electrons obviously increases the 
contribution from the two-body interaction. Since our optimal
coupled-cluster results are very close to the DMC results, almost
identical results are obtained if we replace the DMC results with 
the $\Lambda-$CCSD(T) results.

We can also study the role of correlations beyond the Hartree-Fock 
energy $E_0$. In order to do this, we relate the Hartree-Fock energy 
$E_0$ in Tables \ref{table:tab2}-\ref{table:tab4} to the optimal 
coupled-cluster calculation, namely the $\Lambda$-CCSD(T) results. 
The relative difference between these quantities conveys thereby 
information about correlations beyond the mean-field approximation.  
This relative measure is defined as
\begin{equation}
\chi = \left|\frac{E_{\Lambda-\mathrm{CCSD(T)}}-E_0}{E_{\Lambda-\mathrm{CCSD(T)}}}\right|.
\label{eq:cce0}
\end{equation}
The results are shown in Fig.~\ref{fig:fig2} for $N=6$, $N=12$, and $N=20$.  

We see from this figure that correlations beyond the Hartree-Fock
level are important for few particles and low values of $\omega$. 
Increasing the number of electrons in the circular dot decreases the 
role of correlations beyond the mean-field approximation, a feature which can be 
understood from the fact that for larger systems, multi-particle 
excitations across the Fermi level  decrease in importance. This is 
due to the fact that  the single-particle wave functions for many 
 states around the Fermi level have more than one node, resulting in 
normally smaller matrix elements. Stated differently, with an increasing 
number of electrons, the particles close to the Fermi level 
are more apart from each other, in particular for those particles
which occupy states around and above the Fermi level. The consequence 
of this is that correlations beyond the Hartree-Fock level decrease in 
importance when we add more and more particles. This means in turn
that for larger systems, mean-field methods are rather good
approximations to systems of many interacting electrons in quantum dots.
Similar features are seen in nuclei. For light nuclei, correlations 
beyond the mean field are very important for ground-state properties, 
whereas for heavy nuclei like $^{208}$Pb mean-field approaches provide 
a very good starting point for studying several observables. 

The reader should, however, note that here we have limited our attention to 
ground-state energies only. Whether our conclusions about the role of
correlations pertain to quantities like say spectroscopic factors 
remains to be studied.

We conclude this section by studying in more detail the convergence properties of
our coupled-cluster approaches, in particular, we will relate our  $\Lambda$-CCSD(T)
and CCSD results with the diffusion Monte Carlo results and study the dependence on $R$.
This analysis will be performed with and without an effective Coulomb interaction.  
The reason for doing this is that we wish to study whether the convergence criterion of Eq.~(\ref{eq:simenproof}), derived for a full configuration interaction analysis, applies to various coupled-cluster truncations as well. Furthermore,
we wish to see whether our calculations with  an effective interaction converge faster as a 
function  of $R$ compared to  a calculation with the bare interaction.   

We compute the following quantities
\begin{equation}
\log_{10}{\epsilon_{\mathrm{CCSD}}(R)} = \log_{10}{\left|\frac{E_{\mathrm{CCSD}}(R)-E_{\mathrm{DMC}}}{E_{\mathrm{DMC}}}\right|},
\label{eq:cce1}
\end{equation}
and 
\begin{equation}
\log_{10}{\epsilon_{\mathrm{\Lambda-CCSD(T)}}(R)} = \log_{10}{\left|\frac{E_{\mathrm{\Lambda-CCSD(T)}}(R)-E_{\mathrm{DMC}}}{E_{\mathrm{DMC}}}\right|}.
\label{eq:cce2}
\end{equation}

In Fig.~\ref{fig:fig3} we plot the results for $N=20$ electrons and $\omega=0.5$.  
We have chosen these values since they represent one of the cases where the 
$\Lambda$-CCSD(T) results are always above the DMC results and we have no crossing between these two sets of calculations. The CCSD results on the other hand are always, for all cases reported here, above the DMC results. This means that the trend seen in Fig.~\ref{fig:fig3}
for the CCSD calculations applies to all cases listed in Tables \ref{table:tab2}-\ref{table:tab4} while for the $\Lambda$-CCSD(T) calculations, these results are similar for all cases except for $N=6$ and $\omega=0.5$ and $\omega=1.0$, $N=12$, and $\omega=1.0$, and $N=20$ and $\omega=1.0$. In these cases the results at $R=20$ are slightly below the DMC results. However, the agreement is still excellent. 

The interesting feature to note in Fig.~\ref{fig:fig3} is that the CCSD results change marginally after $R=12$ for $N=20$, and there is essentially very little to gain beyond twenty major shells.
With the present accuracy of the DMC results, we can conclude that the CCSD results reach at most a 
relative error of approximately $10^{-3}$ and that it stays almost stable from $R=12$ shells.
The relative error with respect to the Monte Carlo results does not change much. This applies to all CCSD results. This tells us clearly that there are important correlations beyond 
two-particle-two-hole excitations and that these correlations do not stabilize after some few shells. 
Furthermore, the slope of the 
$\Lambda$-CCSD(T) calculations is much more interesting and resembles the slope of the configuration interaction analysis of Ref.~\onlinecite{kvaal2009} with an effective interaction. 
For the ground states of three to five electrons, Kvaal found in Ref.~\onlinecite{kvaal2009}
a slope of approximately $\alpha=-4$ to $\alpha=-5$ for a parameterization 
\[
\log_{10}{\epsilon}\approx c+\alpha \log_{10}{R},
\]
for the ground-state energies of various $N$-electron quantum dots. The variable 
$c$ is a constant. Our slopes vary between $\alpha=-4$ and $\alpha=-6$, resulting in a relative error
of approximately $10^{-5}$ at  $R=20$ for the results in Fig.~\ref{fig:fig3}. The slope of the 
$\Lambda$-CCSD(T) result is $\alpha = -4.93$. The reader should  note
that the DMC results cannot reach a higher precision. The slope of the CCSD calculation with an effective
interaction is $\alpha=-0.67$ after $R=12$. 

In the same figure, we plot also the $\Lambda$-CCSD(T) results obtained without an effective Coulomb interaction, that is, with the bare interaction only. These results are labelled as 
$\log_{10}{\epsilon_{\mathrm{\Lambda-CCSD(T)}}(R)}-$bare. A Hartree-Fock basis was used in this case as well in order
to obtain converged solutions for the $\Lambda$-CCSD(T) equations. We see in this case that the convergence is much slower, resulting in a slope given by $\alpha = -2.58$, a result not far from the analysis of Ref.~\onlinecite{kvaal2009} for the bare interaction. 
Figure \ref{fig:fig4} exhibits a similar trend, except that here we present results  for $N=12$ electrons 
and $\omega=0.5$. The slope of the 
$\Lambda$-CCSD(T) results is now $\alpha = -6.38$ with an effective interaction and $\alpha=-1.81$ with a bare Coulomb 
interaction. We notice again that the CCSD results saturate around $R=12$ major shells.
 
These results are very interesting as they show that the usage of an effective interaction can really speed up
the convergence of the energy as a function of the number of shells. Furthermore, these results  tell us also that correlations beyond the singles and doubles approach are simply necessary.  
The convergence behavior of the $\Lambda$-CCSD(T) results resembles, to a large extent, those of a
full configuration interaction approach with and without an effective interaction. 
Although we can extract similar convergence behaviors as those 
predicted in Ref.~\onlinecite{kvaal2009} as functions of a truncation in the single-particle basis,  
the challenge is to provide more rigid mathematical 
convergence criteria for truncations in the number of particle-hole excitations. Here we can only justify {\em a posteriori} that triples corrections are necessary. Work along these lines is in progress.

\section{Conclusions and perspectives}

We have shown in this work that 
coupled-cluster calculations that employ an effective Coulomb 
interaction and a self-consistent Hartree-Fock single-particle basis 
reproduce excellently diffusion Monte Carlo calculations, 
even for very low oscillator frequencies. This opens up many
interesting perspectives, in particular since our coupled-cluster 
calculations are rather inexpensive from a high-performance computing 
standpoint. Properties like addition spectra and excited states can be 
extracted using equation-of-motion-based techniques (see, for example, 
Refs.~\onlinecite{bartlett2007,hagen2010a,hagen2010b}). Furthermore, since
our codes run in an uncoupled basis, one can also study other trapping 
potentials than the standard harmonic oscillator potential. A
time-dependent formulation of coupled-cluster theory may even allow 
for studies of temporal properties of quantum dots such as the effect 
of a time-dependent perturbation. 

For circular dots, we found that with the inclusion of triples
correlations, there are, for all systems studied, indications 
that many-body correlations beyond three-particle-three-hole 
excitations in the coupled-cluster amplitude $T$, are negligible. 
We observe also that for systems with more particles, correlations beyond 
the Hartree-Fock level tend to decrease. Thus, although we are 
able to extend {\em ab initio} coupled-cluster calculations of 
quantum dots to systems up to $50$ electrons, a mean-field description 
will probably convey most of the interesting physics.  

With two popular and reliable many-body techniques like coupled-cluster 
theory and diffusion Monte Carlo resulting in practically 
the same energies, one is in the position where one can extract almost 
exact density functionals for quantum dot systems. This allows for 
important comparisons with available density functionals for quantum dots.    Finally, we have also noted that triples correlations are necessary in order to obtain correct results. The convergence
pattern of our calculations resemble to a large extent those seen in full configuration
interaction calculations.

This work was supported by the Research Council of Norway, 
by the Office of Nuclear Physics, U.S. Department of Energy
(Oak Ridge National Laboratory); the University of
Washington under Contract No. DE-FC02-07ER41457. 
This research used computational resources of the
National Center for Computational Sciences at Oak Ridge National Laboratory, 
the supercomputing center Titan at the University of Oslo and the supercomputing
center at the University of Trento.

\begin{widetext}
\begin{figure}
\input{fig1.tex}
\caption{Relative correlation energy $\epsilon$ defined in Eq.~(\ref{eq:dmch0}) 
for different values of $\hbar\omega$ and number of electrons. The DMC numbers are obtained from Tables \ref{table:tab1}, and \ref{table:tab2}-\ref{table:tab4} using $R=20$.}
\label{fig:fig1}
\end{figure}
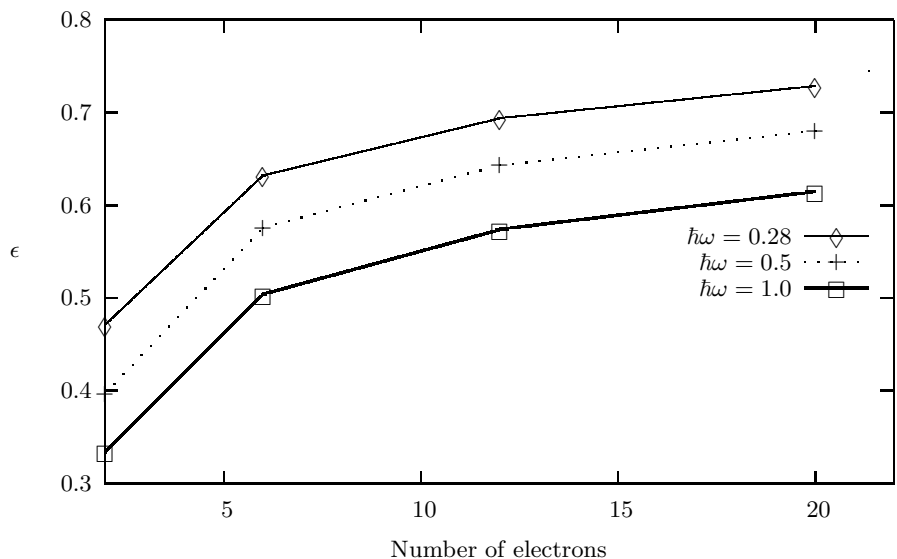
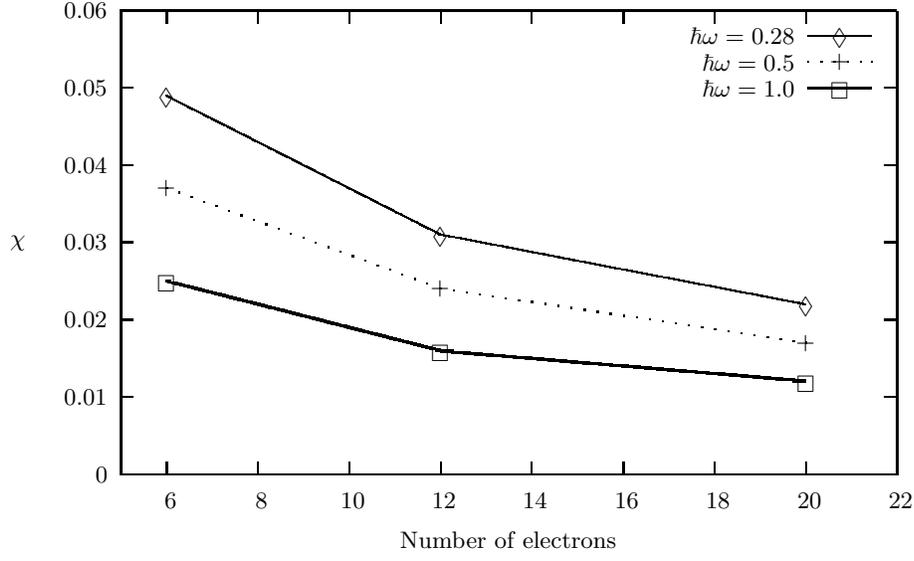
\begin{figure}
\input{fig2.tex}
\caption{Relative correlation energy $\chi$ defined in Eq.~(\ref{eq:cce0}) 
for different values of $\hbar\omega$ and number of electrons. The numbers are obtained 
from Tables \ref{table:tab2}-\ref{table:tab4} using $R=20$.}
\label{fig:fig2}
\end{figure}
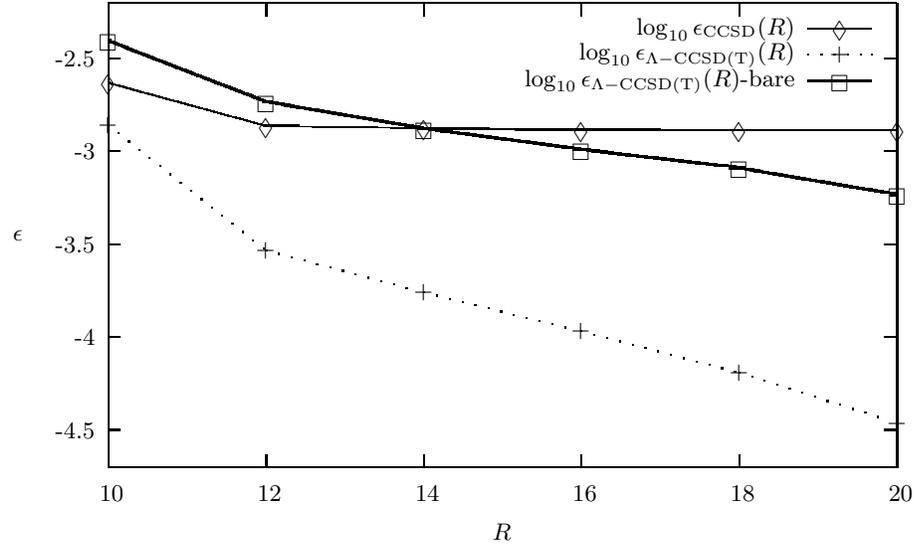
\begin{figure}
\input{fig3.tex}
\caption{Relative correlation energy $\epsilon$ defined in Eqs.~(\ref{eq:cce1}) and (\ref{eq:cce2}) 
for different values of $R$. The values displayed here are for $N=20$ and $\omega=0.5$. The numbers are obtained 
from Table \ref{table:tab4}. We include also the $\Lambda$-CCSD(T) results obtained with the bare interaction.}
\label{fig:fig3}
\end{figure}
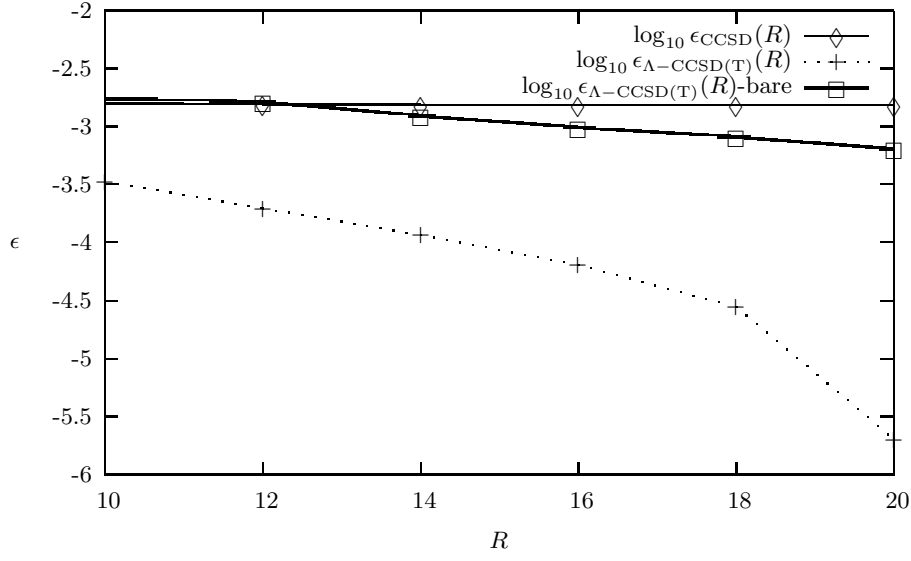
\begin{figure}
\input{fig4.tex}
\caption{Relative correlation energy $\epsilon$ defined in Eqs.~(\ref{eq:cce1}) and (\ref{eq:cce2}) 
for different values of $R$. The values displayed here are for $N=12$ and $\omega=0.5$. The numbers are obtained 
from Table \ref{table:tab3}. We include also the $\Lambda$-CCSD(T) results obtained with the bare interaction.}
\label{fig:fig4}
\end{figure}

\end{widetext}

\end{document}

%% file: fig1.tex
\setlength{\unitlength}{0.240900pt}
\ifx\plotpoint\undefined\newsavebox{\plotpoint}\fi
\sbox{\plotpoint}{\rule[-0.200pt]{0.400pt}{0.400pt}}%
\begin{picture}(1500,900)(0,0)
\sbox{\plotpoint}{\rule[-0.200pt]{0.400pt}{0.400pt}}%
\put(211.0,131.0){\rule[-0.200pt]{4.818pt}{0.400pt}}
\put(191,131){\makebox(0,0)[r]{ 0.3}}
\put(1430.0,131.0){\rule[-0.200pt]{4.818pt}{0.400pt}}
\put(211.0,277.0){\rule[-0.200pt]{4.818pt}{0.400pt}}
\put(191,277){\makebox(0,0)[r]{ 0.4}}
\put(1430.0,277.0){\rule[-0.200pt]{4.818pt}{0.400pt}}
\put(211.0,423.0){\rule[-0.200pt]{4.818pt}{0.400pt}}
\put(191,423){\makebox(0,0)[r]{ 0.5}}
\put(1430.0,423.0){\rule[-0.200pt]{4.818pt}{0.400pt}}
\put(211.0,568.0){\rule[-0.200pt]{4.818pt}{0.400pt}}
\put(191,568){\makebox(0,0)[r]{ 0.6}}
\put(1430.0,568.0){\rule[-0.200pt]{4.818pt}{0.400pt}}
\put(211.0,714.0){\rule[-0.200pt]{4.818pt}{0.400pt}}
\put(191,714){\makebox(0,0)[r]{ 0.7}}
\put(1430.0,714.0){\rule[-0.200pt]{4.818pt}{0.400pt}}
\put(211.0,860.0){\rule[-0.200pt]{4.818pt}{0.400pt}}
\put(191,860){\makebox(0,0)[r]{ 0.8}}
\put(1430.0,860.0){\rule[-0.200pt]{4.818pt}{0.400pt}}
\put(397.0,131.0){\rule[-0.200pt]{0.400pt}{4.818pt}}
\put(397,90){\makebox(0,0){ 5}}
\put(397.0,840.0){\rule[-0.200pt]{0.400pt}{4.818pt}}
\put(707.0,131.0){\rule[-0.200pt]{0.400pt}{4.818pt}}
\put(707,90){\makebox(0,0){ 10}}
\put(707.0,840.0){\rule[-0.200pt]{0.400pt}{4.818pt}}
\put(1016.0,131.0){\rule[-0.200pt]{0.400pt}{4.818pt}}
\put(1016,90){\makebox(0,0){ 15}}
\put(1016.0,840.0){\rule[-0.200pt]{0.400pt}{4.818pt}}
\put(1326.0,131.0){\rule[-0.200pt]{0.400pt}{4.818pt}}
\put(1326,90){\makebox(0,0){ 20}}
\put(1326.0,840.0){\rule[-0.200pt]{0.400pt}{4.818pt}}
\put(211.0,131.0){\rule[-0.200pt]{0.400pt}{175.616pt}}
\put(211.0,131.0){\rule[-0.200pt]{298.475pt}{0.400pt}}
\put(1450.0,131.0){\rule[-0.200pt]{0.400pt}{175.616pt}}
\put(211.0,860.0){\rule[-0.200pt]{298.475pt}{0.400pt}}
\put(70,495){\makebox(0,0){$\epsilon$}}
\put(830,29){\makebox(0,0){Number of electrons}}
\put(1290,520){\makebox(0,0)[r]{$\hbar\omega = 0.28$}}
\put(1310.0,520.0){\rule[-0.200pt]{24.090pt}{0.400pt}}
\put(211,380){\usebox{\plotpoint}}
\multiput(211.00,380.58)(0.528,0.500){467}{\rule{0.522pt}{0.120pt}}
\multiput(211.00,379.17)(246.916,235.000){2}{\rule{0.261pt}{0.400pt}}
\multiput(459.00,615.58)(2.072,0.499){177}{\rule{1.753pt}{0.120pt}}
\multiput(459.00,614.17)(368.361,90.000){2}{\rule{0.877pt}{0.400pt}}
\multiput(831.00,705.58)(4.882,0.498){99}{\rule{3.982pt}{0.120pt}}
\multiput(831.00,704.17)(486.734,51.000){2}{\rule{1.991pt}{0.400pt}}
\put(211,380){\raisebox{-.8pt}{\makebox(0,0){$\Diamond$}}}
\put(459,615){\raisebox{-.8pt}{\makebox(0,0){$\Diamond$}}}
\put(831,705){\raisebox{-.8pt}{\makebox(0,0){$\Diamond$}}}
\put(1326,756){\raisebox{-.8pt}{\makebox(0,0){$\Diamond$}}}
\put(1360,520){\raisebox{-.8pt}{\makebox(0,0){$\Diamond$}}}
\put(1290,479){\makebox(0,0)[r]{$\hbar\omega = 0.5$}}
\multiput(1310,479)(20.756,0.000){5}{\usebox{\plotpoint}}
\put(1410,779){\usebox{\plotpoint}}
\put(211,272){\usebox{\plotpoint}}
\multiput(211,272)(14.297,15.046){18}{\usebox{\plotpoint}}
\multiput(459,533)(20.071,5.287){18}{\usebox{\plotpoint}}
\multiput(831,631)(20.633,2.251){24}{\usebox{\plotpoint}}
\put(1326,685){\usebox{\plotpoint}}
\put(211,272){\makebox(0,0){$+$}}
\put(459,533){\makebox(0,0){$+$}}
\put(831,631){\makebox(0,0){$+$}}
\put(1326,685){\makebox(0,0){$+$}}
\put(1360,479){\makebox(0,0){$+$}}
\sbox{\plotpoint}{\rule[-0.400pt]{0.800pt}{0.800pt}}%
\sbox{\plotpoint}{\rule[-0.200pt]{0.400pt}{0.400pt}}%
\put(1290,438){\makebox(0,0)[r]{$\hbar\omega = 1.0$}}
\sbox{\plotpoint}{\rule[-0.400pt]{0.800pt}{0.800pt}}%
\put(1310.0,438.0){\rule[-0.400pt]{24.090pt}{0.800pt}}
\put(211,180){\usebox{\plotpoint}}
\multiput(211.00,181.41)(0.500,0.500){489}{\rule{1.000pt}{0.121pt}}
\multiput(211.00,178.34)(245.924,248.000){2}{\rule{0.500pt}{0.800pt}}
\multiput(459.00,429.41)(1.832,0.501){197}{\rule{3.118pt}{0.121pt}}
\multiput(459.00,426.34)(365.529,102.000){2}{\rule{1.559pt}{0.800pt}}
\multiput(831.00,531.41)(4.168,0.502){113}{\rule{6.800pt}{0.121pt}}
\multiput(831.00,528.34)(480.886,60.000){2}{\rule{3.400pt}{0.800pt}}
\put(211,180){\raisebox{-.8pt}{\makebox(0,0){$\Box$}}}
\put(459,428){\raisebox{-.8pt}{\makebox(0,0){$\Box$}}}
\put(831,530){\raisebox{-.8pt}{\makebox(0,0){$\Box$}}}
\put(1326,590){\raisebox{-.8pt}{\makebox(0,0){$\Box$}}}
\put(1360,438){\raisebox{-.8pt}{\makebox(0,0){$\Box$}}}
\sbox{\plotpoint}{\rule[-0.200pt]{0.400pt}{0.400pt}}%
\put(211.0,131.0){\rule[-0.200pt]{0.400pt}{175.616pt}}
\put(211.0,131.0){\rule[-0.200pt]{298.475pt}{0.400pt}}
\put(1450.0,131.0){\rule[-0.200pt]{0.400pt}{175.616pt}}
\put(211.0,860.0){\rule[-0.200pt]{298.475pt}{0.400pt}}
\end{picture}

%% file: fig2.tex
\setlength{\unitlength}{0.240900pt}
\ifx\plotpoint\undefined\newsavebox{\plotpoint}\fi
\sbox{\plotpoint}{\rule[-0.200pt]{0.400pt}{0.400pt}}%
\begin{picture}(1500,900)(0,0)
\sbox{\plotpoint}{\rule[-0.200pt]{0.400pt}{0.400pt}}%
\put(231.0,131.0){\rule[-0.200pt]{4.818pt}{0.400pt}}
\put(211,131){\makebox(0,0)[r]{ 0}}
\put(1430.0,131.0){\rule[-0.200pt]{4.818pt}{0.400pt}}
\put(231.0,253.0){\rule[-0.200pt]{4.818pt}{0.400pt}}
\put(211,253){\makebox(0,0)[r]{ 0.01}}
\put(1430.0,253.0){\rule[-0.200pt]{4.818pt}{0.400pt}}
\put(231.0,374.0){\rule[-0.200pt]{4.818pt}{0.400pt}}
\put(211,374){\makebox(0,0)[r]{ 0.02}}
\put(1430.0,374.0){\rule[-0.200pt]{4.818pt}{0.400pt}}
\put(231.0,496.0){\rule[-0.200pt]{4.818pt}{0.400pt}}
\put(211,496){\makebox(0,0)[r]{ 0.03}}
\put(1430.0,496.0){\rule[-0.200pt]{4.818pt}{0.400pt}}
\put(231.0,617.0){\rule[-0.200pt]{4.818pt}{0.400pt}}
\put(211,617){\makebox(0,0)[r]{ 0.04}}
\put(1430.0,617.0){\rule[-0.200pt]{4.818pt}{0.400pt}}
\put(231.0,739.0){\rule[-0.200pt]{4.818pt}{0.400pt}}
\put(211,739){\makebox(0,0)[r]{ 0.05}}
\put(1430.0,739.0){\rule[-0.200pt]{4.818pt}{0.400pt}}
\put(231.0,860.0){\rule[-0.200pt]{4.818pt}{0.400pt}}
\put(211,860){\makebox(0,0)[r]{ 0.06}}
\put(1430.0,860.0){\rule[-0.200pt]{4.818pt}{0.400pt}}
\put(303.0,131.0){\rule[-0.200pt]{0.400pt}{4.818pt}}
\put(303,90){\makebox(0,0){ 6}}
\put(303.0,840.0){\rule[-0.200pt]{0.400pt}{4.818pt}}
\put(446.0,131.0){\rule[-0.200pt]{0.400pt}{4.818pt}}
\put(446,90){\makebox(0,0){ 8}}
\put(446.0,840.0){\rule[-0.200pt]{0.400pt}{4.818pt}}
\put(590.0,131.0){\rule[-0.200pt]{0.400pt}{4.818pt}}
\put(590,90){\makebox(0,0){ 10}}
\put(590.0,840.0){\rule[-0.200pt]{0.400pt}{4.818pt}}
\put(733.0,131.0){\rule[-0.200pt]{0.400pt}{4.818pt}}
\put(733,90){\makebox(0,0){ 12}}
\put(733.0,840.0){\rule[-0.200pt]{0.400pt}{4.818pt}}
\put(876.0,131.0){\rule[-0.200pt]{0.400pt}{4.818pt}}
\put(876,90){\makebox(0,0){ 14}}
\put(876.0,840.0){\rule[-0.200pt]{0.400pt}{4.818pt}}
\put(1020.0,131.0){\rule[-0.200pt]{0.400pt}{4.818pt}}
\put(1020,90){\makebox(0,0){ 16}}
\put(1020.0,840.0){\rule[-0.200pt]{0.400pt}{4.818pt}}
\put(1163.0,131.0){\rule[-0.200pt]{0.400pt}{4.818pt}}
\put(1163,90){\makebox(0,0){ 18}}
\put(1163.0,840.0){\rule[-0.200pt]{0.400pt}{4.818pt}}
\put(1307.0,131.0){\rule[-0.200pt]{0.400pt}{4.818pt}}
\put(1307,90){\makebox(0,0){ 20}}
\put(1307.0,840.0){\rule[-0.200pt]{0.400pt}{4.818pt}}
\put(1450.0,131.0){\rule[-0.200pt]{0.400pt}{4.818pt}}
\put(1450,90){\makebox(0,0){ 22}}
\put(1450.0,840.0){\rule[-0.200pt]{0.400pt}{4.818pt}}
\put(231.0,131.0){\rule[-0.200pt]{0.400pt}{175.616pt}}
\put(231.0,131.0){\rule[-0.200pt]{293.657pt}{0.400pt}}
\put(1450.0,131.0){\rule[-0.200pt]{0.400pt}{175.616pt}}
\put(231.0,860.0){\rule[-0.200pt]{293.657pt}{0.400pt}}
\put(70,495){\makebox(0,0){$\chi$}}
\put(840,29){\makebox(0,0){Number of electrons}}
\put(1290,820){\makebox(0,0)[r]{$\hbar\omega = 0.28$}}
\put(1310.0,820.0){\rule[-0.200pt]{24.090pt}{0.400pt}}
\put(303,726){\usebox{\plotpoint}}
\multiput(303.00,724.92)(0.987,-0.500){433}{\rule{0.889pt}{0.120pt}}
\multiput(303.00,725.17)(428.155,-218.000){2}{\rule{0.444pt}{0.400pt}}
\multiput(733.00,506.92)(2.615,-0.499){217}{\rule{2.187pt}{0.120pt}}
\multiput(733.00,507.17)(569.460,-110.000){2}{\rule{1.094pt}{0.400pt}}
\put(303,726){\raisebox{-.8pt}{\makebox(0,0){$\Diamond$}}}
\put(733,508){\raisebox{-.8pt}{\makebox(0,0){$\Diamond$}}}
\put(1307,398){\raisebox{-.8pt}{\makebox(0,0){$\Diamond$}}}
\put(1360,820){\raisebox{-.8pt}{\makebox(0,0){$\Diamond$}}}
\put(1290,779){\makebox(0,0)[r]{$\hbar\omega = 0.5$}}
\multiput(1310,779)(20.756,0.000){5}{\usebox{\plotpoint}}
\put(1410,779){\usebox{\plotpoint}}
\put(303,581){\usebox{\plotpoint}}
\multiput(303,581)(19.482,-7.158){23}{\usebox{\plotpoint}}
\multiput(733,423)(20.532,-3.040){28}{\usebox{\plotpoint}}
\put(1307,338){\usebox{\plotpoint}}
\put(303,581){\makebox(0,0){$+$}}
\put(733,423){\makebox(0,0){$+$}}
\put(1307,338){\makebox(0,0){$+$}}
\put(1360,779){\makebox(0,0){$+$}}
\sbox{\plotpoint}{\rule[-0.400pt]{0.800pt}{0.800pt}}%
\sbox{\plotpoint}{\rule[-0.200pt]{0.400pt}{0.400pt}}%
\put(1290,738){\makebox(0,0)[r]{$\hbar\omega = 1.0$}}
\sbox{\plotpoint}{\rule[-0.400pt]{0.800pt}{0.800pt}}%
\put(1310.0,738.0){\rule[-0.400pt]{24.090pt}{0.800pt}}
\put(303,435){\usebox{\plotpoint}}
\multiput(303.00,433.09)(1.963,-0.501){213}{\rule{3.327pt}{0.121pt}}
\multiput(303.00,433.34)(423.094,-110.000){2}{\rule{1.664pt}{0.800pt}}
\multiput(733.00,323.09)(6.062,-0.502){89}{\rule{9.767pt}{0.121pt}}
\multiput(733.00,323.34)(553.729,-48.000){2}{\rule{4.883pt}{0.800pt}}
\put(303,435){\raisebox{-.8pt}{\makebox(0,0){$\Box$}}}
\put(733,325){\raisebox{-.8pt}{\makebox(0,0){$\Box$}}}
\put(1307,277){\raisebox{-.8pt}{\makebox(0,0){$\Box$}}}
\put(1360,738){\raisebox{-.8pt}{\makebox(0,0){$\Box$}}}
\sbox{\plotpoint}{\rule[-0.200pt]{0.400pt}{0.400pt}}%
\put(231.0,131.0){\rule[-0.200pt]{0.400pt}{175.616pt}}
\put(231.0,131.0){\rule[-0.200pt]{293.657pt}{0.400pt}}
\put(1450.0,131.0){\rule[-0.200pt]{0.400pt}{175.616pt}}
\put(231.0,860.0){\rule[-0.200pt]{293.657pt}{0.400pt}}
\end{picture}

%% file: fig3.tex
\setlength{\unitlength}{0.240900pt}
\ifx\plotpoint\undefined\newsavebox{\plotpoint}\fi
\sbox{\plotpoint}{\rule[-0.200pt]{0.400pt}{0.400pt}}%
\begin{picture}(1500,900)(0,0)
\sbox{\plotpoint}{\rule[-0.200pt]{0.400pt}{0.400pt}}%
\put(211.0,189.0){\rule[-0.200pt]{4.818pt}{0.400pt}}
\put(191,189){\makebox(0,0)[r]{-4.5}}
\put(1430.0,189.0){\rule[-0.200pt]{4.818pt}{0.400pt}}
\put(211.0,335.0){\rule[-0.200pt]{4.818pt}{0.400pt}}
\put(191,335){\makebox(0,0)[r]{-4}}
\put(1430.0,335.0){\rule[-0.200pt]{4.818pt}{0.400pt}}
\put(211.0,481.0){\rule[-0.200pt]{4.818pt}{0.400pt}}
\put(191,481){\makebox(0,0)[r]{-3.5}}
\put(1430.0,481.0){\rule[-0.200pt]{4.818pt}{0.400pt}}
\put(211.0,627.0){\rule[-0.200pt]{4.818pt}{0.400pt}}
\put(191,627){\makebox(0,0)[r]{-3}}
\put(1430.0,627.0){\rule[-0.200pt]{4.818pt}{0.400pt}}
\put(211.0,773.0){\rule[-0.200pt]{4.818pt}{0.400pt}}
\put(191,773){\makebox(0,0)[r]{-2.5}}
\put(1430.0,773.0){\rule[-0.200pt]{4.818pt}{0.400pt}}
\put(211.0,131.0){\rule[-0.200pt]{0.400pt}{4.818pt}}
\put(211,90){\makebox(0,0){ 10}}
\put(211.0,840.0){\rule[-0.200pt]{0.400pt}{4.818pt}}
\put(459.0,131.0){\rule[-0.200pt]{0.400pt}{4.818pt}}
\put(459,90){\makebox(0,0){ 12}}
\put(459.0,840.0){\rule[-0.200pt]{0.400pt}{4.818pt}}
\put(707.0,131.0){\rule[-0.200pt]{0.400pt}{4.818pt}}
\put(707,90){\makebox(0,0){ 14}}
\put(707.0,840.0){\rule[-0.200pt]{0.400pt}{4.818pt}}
\put(954.0,131.0){\rule[-0.200pt]{0.400pt}{4.818pt}}
\put(954,90){\makebox(0,0){ 16}}
\put(954.0,840.0){\rule[-0.200pt]{0.400pt}{4.818pt}}
\put(1202.0,131.0){\rule[-0.200pt]{0.400pt}{4.818pt}}
\put(1202,90){\makebox(0,0){ 18}}
\put(1202.0,840.0){\rule[-0.200pt]{0.400pt}{4.818pt}}
\put(1450.0,131.0){\rule[-0.200pt]{0.400pt}{4.818pt}}
\put(1450,90){\makebox(0,0){ 20}}
\put(1450.0,840.0){\rule[-0.200pt]{0.400pt}{4.818pt}}
\put(211.0,131.0){\rule[-0.200pt]{0.400pt}{175.616pt}}
\put(211.0,131.0){\rule[-0.200pt]{298.475pt}{0.400pt}}
\put(1450.0,131.0){\rule[-0.200pt]{0.400pt}{175.616pt}}
\put(211.0,860.0){\rule[-0.200pt]{298.475pt}{0.400pt}}
\put(70,495){\makebox(0,0){$\epsilon$}}
\put(830,29){\makebox(0,0){$R$}}
\put(1290,820){\makebox(0,0)[r]{$\log_{10}{\epsilon_{\mathrm{CCSD}}(R)}$}}
\put(1310.0,820.0){\rule[-0.200pt]{24.090pt}{0.400pt}}
\put(211,735){\usebox{\plotpoint}}
\multiput(211.00,733.92)(1.830,-0.499){133}{\rule{1.559pt}{0.120pt}}
\multiput(211.00,734.17)(244.765,-68.000){2}{\rule{0.779pt}{0.400pt}}
\multiput(459.00,665.94)(36.159,-0.468){5}{\rule{24.900pt}{0.113pt}}
\multiput(459.00,666.17)(196.319,-4.000){2}{\rule{12.450pt}{0.400pt}}
\put(707,661.17){\rule{49.500pt}{0.400pt}}
\multiput(707.00,662.17)(144.260,-2.000){2}{\rule{24.750pt}{0.400pt}}
\put(954,659.67){\rule{59.743pt}{0.400pt}}
\multiput(954.00,660.17)(124.000,-1.000){2}{\rule{29.872pt}{0.400pt}}
\put(211,735){\raisebox{-.8pt}{\makebox(0,0){$\Diamond$}}}
\put(459,667){\raisebox{-.8pt}{\makebox(0,0){$\Diamond$}}}
\put(707,663){\raisebox{-.8pt}{\makebox(0,0){$\Diamond$}}}
\put(954,661){\raisebox{-.8pt}{\makebox(0,0){$\Diamond$}}}
\put(1202,660){\raisebox{-.8pt}{\makebox(0,0){$\Diamond$}}}
\put(1450,660){\raisebox{-.8pt}{\makebox(0,0){$\Diamond$}}}
\put(1360,820){\raisebox{-.8pt}{\makebox(0,0){$\Diamond$}}}
\put(1202.0,660.0){\rule[-0.200pt]{59.743pt}{0.400pt}}
\put(1290,779){\makebox(0,0)[r]{$\log_{10}{\epsilon_{\mathrm{\Lambda-CCSD(T)}}(R)}$}}
\multiput(1310,779)(20.756,0.000){5}{\usebox{\plotpoint}}
\put(1410,779){\usebox{\plotpoint}}
\put(211,668){\usebox{\plotpoint}}
\multiput(211,668)(16.252,-12.910){16}{\usebox{\plotpoint}}
\multiput(459,471)(20.057,-5.338){12}{\usebox{\plotpoint}}
\multiput(707,405)(20.150,-4.976){12}{\usebox{\plotpoint}}
\multiput(954,344)(20.077,-5.262){13}{\usebox{\plotpoint}}
\multiput(1202,279)(19.776,-6.300){12}{\usebox{\plotpoint}}
\put(1450,200){\usebox{\plotpoint}}
\put(211,668){\makebox(0,0){$+$}}
\put(459,471){\makebox(0,0){$+$}}
\put(707,405){\makebox(0,0){$+$}}
\put(954,344){\makebox(0,0){$+$}}
\put(1202,279){\makebox(0,0){$+$}}
\put(1450,200){\makebox(0,0){$+$}}
\put(1360,779){\makebox(0,0){$+$}}
\sbox{\plotpoint}{\rule[-0.400pt]{0.800pt}{0.800pt}}%
\sbox{\plotpoint}{\rule[-0.200pt]{0.400pt}{0.400pt}}%
\put(1290,738){\makebox(0,0)[r]{$\log_{10}{\epsilon_{\mathrm{\Lambda-CCSD(T)}}(R)}$-bare}}
\sbox{\plotpoint}{\rule[-0.400pt]{0.800pt}{0.800pt}}%
\put(1310.0,738.0){\rule[-0.400pt]{24.090pt}{0.800pt}}
\put(211,801){\usebox{\plotpoint}}
\multiput(211.00,799.09)(1.297,-0.501){185}{\rule{2.267pt}{0.121pt}}
\multiput(211.00,799.34)(243.295,-96.000){2}{\rule{1.133pt}{0.800pt}}
\multiput(459.00,703.09)(2.994,-0.502){77}{\rule{4.924pt}{0.121pt}}
\multiput(459.00,703.34)(237.780,-42.000){2}{\rule{2.462pt}{0.800pt}}
\multiput(707.00,661.09)(3.816,-0.503){59}{\rule{6.188pt}{0.121pt}}
\multiput(707.00,661.34)(234.157,-33.000){2}{\rule{3.094pt}{0.800pt}}
\multiput(954.00,628.09)(4.375,-0.504){51}{\rule{7.041pt}{0.121pt}}
\multiput(954.00,628.34)(233.385,-29.000){2}{\rule{3.521pt}{0.800pt}}
\multiput(1202.00,599.09)(2.994,-0.502){77}{\rule{4.924pt}{0.121pt}}
\multiput(1202.00,599.34)(237.780,-42.000){2}{\rule{2.462pt}{0.800pt}}
\put(211,801){\raisebox{-.8pt}{\makebox(0,0){$\Box$}}}
\put(459,705){\raisebox{-.8pt}{\makebox(0,0){$\Box$}}}
\put(707,663){\raisebox{-.8pt}{\makebox(0,0){$\Box$}}}
\put(954,630){\raisebox{-.8pt}{\makebox(0,0){$\Box$}}}
\put(1202,601){\raisebox{-.8pt}{\makebox(0,0){$\Box$}}}
\put(1450,559){\raisebox{-.8pt}{\makebox(0,0){$\Box$}}}
\put(1360,738){\raisebox{-.8pt}{\makebox(0,0){$\Box$}}}
\sbox{\plotpoint}{\rule[-0.200pt]{0.400pt}{0.400pt}}%
\put(211.0,131.0){\rule[-0.200pt]{0.400pt}{175.616pt}}
\put(211.0,131.0){\rule[-0.200pt]{298.475pt}{0.400pt}}
\put(1450.0,131.0){\rule[-0.200pt]{0.400pt}{175.616pt}}
\put(211.0,860.0){\rule[-0.200pt]{298.475pt}{0.400pt}}
\end{picture}

%% file: fig4.tex
\setlength{\unitlength}{0.240900pt}
\ifx\plotpoint\undefined\newsavebox{\plotpoint}\fi
\sbox{\plotpoint}{\rule[-0.200pt]{0.400pt}{0.400pt}}%
\begin{picture}(1500,900)(0,0)
\sbox{\plotpoint}{\rule[-0.200pt]{0.400pt}{0.400pt}}%
\put(211.0,131.0){\rule[-0.200pt]{4.818pt}{0.400pt}}
\put(191,131){\makebox(0,0)[r]{-6}}
\put(1430.0,131.0){\rule[-0.200pt]{4.818pt}{0.400pt}}
\put(211.0,222.0){\rule[-0.200pt]{4.818pt}{0.400pt}}
\put(191,222){\makebox(0,0)[r]{-5.5}}
\put(1430.0,222.0){\rule[-0.200pt]{4.818pt}{0.400pt}}
\put(211.0,313.0){\rule[-0.200pt]{4.818pt}{0.400pt}}
\put(191,313){\makebox(0,0)[r]{-5}}
\put(1430.0,313.0){\rule[-0.200pt]{4.818pt}{0.400pt}}
\put(211.0,404.0){\rule[-0.200pt]{4.818pt}{0.400pt}}
\put(191,404){\makebox(0,0)[r]{-4.5}}
\put(1430.0,404.0){\rule[-0.200pt]{4.818pt}{0.400pt}}
\put(211.0,496.0){\rule[-0.200pt]{4.818pt}{0.400pt}}
\put(191,496){\makebox(0,0)[r]{-4}}
\put(1430.0,496.0){\rule[-0.200pt]{4.818pt}{0.400pt}}
\put(211.0,587.0){\rule[-0.200pt]{4.818pt}{0.400pt}}
\put(191,587){\makebox(0,0)[r]{-3.5}}
\put(1430.0,587.0){\rule[-0.200pt]{4.818pt}{0.400pt}}
\put(211.0,678.0){\rule[-0.200pt]{4.818pt}{0.400pt}}
\put(191,678){\makebox(0,0)[r]{-3}}
\put(1430.0,678.0){\rule[-0.200pt]{4.818pt}{0.400pt}}
\put(211.0,769.0){\rule[-0.200pt]{4.818pt}{0.400pt}}
\put(191,769){\makebox(0,0)[r]{-2.5}}
\put(1430.0,769.0){\rule[-0.200pt]{4.818pt}{0.400pt}}
\put(211.0,860.0){\rule[-0.200pt]{4.818pt}{0.400pt}}
\put(191,860){\makebox(0,0)[r]{-2}}
\put(1430.0,860.0){\rule[-0.200pt]{4.818pt}{0.400pt}}
\put(211.0,131.0){\rule[-0.200pt]{0.400pt}{4.818pt}}
\put(211,90){\makebox(0,0){ 10}}
\put(211.0,840.0){\rule[-0.200pt]{0.400pt}{4.818pt}}
\put(459.0,131.0){\rule[-0.200pt]{0.400pt}{4.818pt}}
\put(459,90){\makebox(0,0){ 12}}
\put(459.0,840.0){\rule[-0.200pt]{0.400pt}{4.818pt}}
\put(707.0,131.0){\rule[-0.200pt]{0.400pt}{4.818pt}}
\put(707,90){\makebox(0,0){ 14}}
\put(707.0,840.0){\rule[-0.200pt]{0.400pt}{4.818pt}}
\put(954.0,131.0){\rule[-0.200pt]{0.400pt}{4.818pt}}
\put(954,90){\makebox(0,0){ 16}}
\put(954.0,840.0){\rule[-0.200pt]{0.400pt}{4.818pt}}
\put(1202.0,131.0){\rule[-0.200pt]{0.400pt}{4.818pt}}
\put(1202,90){\makebox(0,0){ 18}}
\put(1202.0,840.0){\rule[-0.200pt]{0.400pt}{4.818pt}}
\put(1450.0,131.0){\rule[-0.200pt]{0.400pt}{4.818pt}}
\put(1450,90){\makebox(0,0){ 20}}
\put(1450.0,840.0){\rule[-0.200pt]{0.400pt}{4.818pt}}
\put(211.0,131.0){\rule[-0.200pt]{0.400pt}{175.616pt}}
\put(211.0,131.0){\rule[-0.200pt]{298.475pt}{0.400pt}}
\put(1450.0,131.0){\rule[-0.200pt]{0.400pt}{175.616pt}}
\put(211.0,860.0){\rule[-0.200pt]{298.475pt}{0.400pt}}
\put(70,495){\makebox(0,0){$\epsilon$}}
\put(830,29){\makebox(0,0){$R$}}
\put(1290,820){\makebox(0,0)[r]{$\log_{10}{\epsilon_{\mathrm{CCSD}}(R)}$}}
\put(211,712.67){\rule{59.743pt}{0.400pt}}
\multiput(211.00,713.17)(124.000,-1.000){2}{\rule{29.872pt}{0.400pt}}
\put(459,711.67){\rule{59.743pt}{0.400pt}}
\multiput(459.00,712.17)(124.000,-1.000){2}{\rule{29.872pt}{0.400pt}}
\put(1310.0,820.0){\rule[-0.200pt]{24.090pt}{0.400pt}}
\put(459,713){\raisebox{-.8pt}{\makebox(0,0){$\Diamond$}}}
\put(707,712){\raisebox{-.8pt}{\makebox(0,0){$\Diamond$}}}
\put(954,712){\raisebox{-.8pt}{\makebox(0,0){$\Diamond$}}}
\put(1202,712){\raisebox{-.8pt}{\makebox(0,0){$\Diamond$}}}
\put(1450,712){\raisebox{-.8pt}{\makebox(0,0){$\Diamond$}}}
\put(1360,820){\raisebox{-.8pt}{\makebox(0,0){$\Diamond$}}}
\put(707.0,712.0){\rule[-0.200pt]{178.989pt}{0.400pt}}
\put(1290,779){\makebox(0,0)[r]{$\log_{10}{\epsilon_{\mathrm{\Lambda-CCSD(T)}}(R)}$}}
\multiput(1310,779)(20.756,0.000){5}{\usebox{\plotpoint}}
\put(1410,779){\usebox{\plotpoint}}
\put(211,591){\usebox{\plotpoint}}
\multiput(211,591)(20.464,-3.466){13}{\usebox{\plotpoint}}
\multiput(459,549)(20.464,-3.466){12}{\usebox{\plotpoint}}
\multiput(707,507)(20.390,-3.880){12}{\usebox{\plotpoint}}
\multiput(954,460)(20.057,-5.338){12}{\usebox{\plotpoint}}
\multiput(1202,394)(15.903,-13.338){16}{\usebox{\plotpoint}}
\put(1450,186){\usebox{\plotpoint}}
\put(211,591){\makebox(0,0){$+$}}
\put(459,549){\makebox(0,0){$+$}}
\put(707,507){\makebox(0,0){$+$}}
\put(954,460){\makebox(0,0){$+$}}
\put(1202,394){\makebox(0,0){$+$}}
\put(1450,186){\makebox(0,0){$+$}}
\put(1360,779){\makebox(0,0){$+$}}
\sbox{\plotpoint}{\rule[-0.400pt]{0.800pt}{0.800pt}}%
\sbox{\plotpoint}{\rule[-0.200pt]{0.400pt}{0.400pt}}%
\put(1290,738){\makebox(0,0)[r]{$\log_{10}{\epsilon_{\mathrm{\Lambda-CCSD(T)}}(R)}$-bare}}
\sbox{\plotpoint}{\rule[-0.400pt]{0.800pt}{0.800pt}}%
\multiput(211.00,719.06)(41.227,-0.560){3}{\rule{39.880pt}{0.135pt}}
\multiput(211.00,719.34)(165.227,-5.000){2}{\rule{19.940pt}{0.800pt}}
\multiput(459.00,714.09)(5.826,-0.505){37}{\rule{9.218pt}{0.122pt}}
\multiput(459.00,714.34)(228.867,-22.000){2}{\rule{4.609pt}{0.800pt}}
\multiput(707.00,692.09)(7.164,-0.506){29}{\rule{11.178pt}{0.122pt}}
\multiput(707.00,692.34)(223.800,-18.000){2}{\rule{5.589pt}{0.800pt}}
\multiput(954.00,674.09)(8.739,-0.508){23}{\rule{13.427pt}{0.122pt}}
\multiput(954.00,674.34)(220.132,-15.000){2}{\rule{6.713pt}{0.800pt}}
\multiput(1202.00,659.09)(6.794,-0.506){31}{\rule{10.642pt}{0.122pt}}
\multiput(1202.00,659.34)(225.912,-19.000){2}{\rule{5.321pt}{0.800pt}}
\put(459,716){\raisebox{-.8pt}{\makebox(0,0){$\Box$}}}
\put(707,694){\raisebox{-.8pt}{\makebox(0,0){$\Box$}}}
\put(954,676){\raisebox{-.8pt}{\makebox(0,0){$\Box$}}}
\put(1202,661){\raisebox{-.8pt}{\makebox(0,0){$\Box$}}}
\put(1450,642){\raisebox{-.8pt}{\makebox(0,0){$\Box$}}}
\put(1360,738){\raisebox{-.8pt}{\makebox(0,0){$\Box$}}}
\put(1310.0,738.0){\rule[-0.400pt]{24.090pt}{0.800pt}}
\sbox{\plotpoint}{\rule[-0.200pt]{0.400pt}{0.400pt}}%
\put(211.0,131.0){\rule[-0.200pt]{0.400pt}{175.616pt}}
\put(211.0,131.0){\rule[-0.200pt]{298.475pt}{0.400pt}}
\put(1450.0,131.0){\rule[-0.200pt]{0.400pt}{175.616pt}}
\put(211.0,860.0){\rule[-0.200pt]{298.475pt}{0.400pt}}
\end{picture}

%% file: paper.bbl
\begin{thebibliography}{200}
\bibitem{Note_DMC}DMC calculations for $N=6$ and $N=12$, with $\omega=0.28$,
have been published in Ref. \onlinecite{pederiva2001}. All the other 
results have been computed for this paper.
\end{thebibliography}
